\nonstopmode

\newcommand{\re}{{\mathrm{Re}\,}}
\newcommand{\eV}{\U{eV}}
\newcommand{\atm}{\U{atm}}
\newcommand{\mul}{\cdot}
\newcommand{\rad}{\U{rad}}
\newcommand{\bohr}{\, a_0}
\newcommand{\Hartree}{\U{E_{\mathit h}}}
\newcommand{\angstrom}{\U{\hbox{\AA}}}
\newcommand{\degree}{{}^{\circ}}
\newcommand{\Celsius}{\,\degree\mathrm C}
\newcommand{\ie}{i.e.{}}
\newcommand{\eg}{e.g.{}}
\newcommand{\etal}{\textit{et al.}}
\newcommand{\U}[1]{\,{\rm{#1}}}
\newcommand{\I}[1]{_{\mathrm{#1}}}
\newcommand{\imag}{{\rm i}}
\newcommand{\euler}{\mathrm e}
\newcommand{\mat}[1]{\hbox{\boldmath{$#1$}\unboldmath}}
\newcommand{\Sum}{\sum\limits}
\newcommand{\Int}{\int\limits}
\newcommand{\transpose}{{}^{\textrm{\scriptsize T}}}
\newcommand{\unitmatrix}{\mat{\mathbbm{1}}}
\newcommand{\differential}{\>\mathrm d}
\newcommand{\bra}[1]{\left<\right.\!#1\!\left.\right|}
\newcommand{\ket}[1]{\left|\right.\!#1\!\left.\right>}
\newcommand{\kilobarns}{\U{kb}}
\newcommand{\E}[1]{\times 10^{#1}}
\newcommand{\dfrac}[2]{{\displaystyle{#1}\over\displaystyle{#2}}}
\renewcommand{\frac}[2]{{#1 \over #2}}

\documentclass[pra,aps,twocolumn,showpacs,nopreprintnumbers,superscriptaddress]{revtex4}
\usepackage{bm,bbm,amssymb,subeqnarray,graphicx}
\usepackage[nativepdf,bookmarks,bookmarksopen,bookmarksnumbered,raiselinks,%
pdftitle={X-ray refractive index of laser-dressed atoms},%
pdfauthor={Christian Buth, Robin Santra},%
pdfsubject={Atomic physics},%
pdfkeywords={quantum optics, ab initio, laser dressing, x rays, %
photoabsorption, cross section, polarization dependence, nonperturbative, %
laser-matter interaction, argon, electromagnetically induced transparency, %
EIT, pulse shaping, index of refraction, refractive index}]{hyperref}

\begin{document}
\title{X-ray refractive index of laser-dressed atoms}
\author{Christian Buth}
\thanks{Present address: Department of Physics and Astronomy, Louisiana State
University, Baton Rouge, Louisiana~70803, USA}
\affiliation{Argonne National Laboratory, Argonne, Illinois~60439, USA}
\author{Robin Santra}
\affiliation{Argonne National Laboratory, Argonne, Illinois~60439, USA}
\affiliation{Department of Physics, University of Chicago, Chicago,
Illinois~60637, USA}
\date{October 14, 2008}

\begin{abstract}
We investigated the complex index of refraction in the x-ray regime of atoms
in laser light.
The laser (intensity up to~$10^{13} \U{W/cm^2}$, wavelength $800 \U{nm}$)
modifies the atomic states but, by assumption, does not excite or ionize
the atoms in their electronic ground state.
Using quantum electrodynamics, we devise an \emph{ab initio} theory to
calculate the dynamic dipole polarizability and the photoabsorption cross
section, which are subsequently used to determine the real and imaginary part,
respectively, of the refractive index.
The interaction with the laser is treated nonperturbatively;
the x-ray interaction is described in terms of a one-photon process.
We numerically solve the resolvents involved using a single-vector Lanczos
algorithm.
Finally, we formulate rate equations to copropagate a laser and an x-ray pulse
through a gas cell.
Our theory is applied to argon.
We study the x-ray polarizability and absorption near the argon
$K$~edge over a large range of dressing-laser intensities.
We find electromagnetically induced transparency~(EIT) for x~rays on the
Ar$\, 1s \to 4p$~pre-edge resonance.
We demonstrate that EIT in Ar allows one to imprint the shape of an
ultrafast laser pulse on a broader x-ray pulse
(duration~$100 \U{ps}$, photon energy~$3.2 \U{keV}$).
Our work thus opens new opportunities for research with hard x-ray sources.
\end{abstract}

%
%
%

\preprint{arXiv:0809.3249}
\maketitle

\section{Introduction}

We study the interaction of atoms with two-color light.
Specifically, we consider an optical laser with moderate intensity
and x~rays.
In experiments this light would be obtained, for instance, from an
amplified Ti:sapphire laser system with up to~$10^{13}\,$W/cm$^2$ at a
wavelength of~$800\U{nm}$.
The x~rays would be produced by a third-generation synchrotron radiation
source like Argonne's Advanced Photon Source~\cite{Thompson:XR-01}.

There are several ways for the chronology between the two light pulses.
First, there is a pump-probe setting where the laser pulse precedes the x~rays.
Especially the combination of a weak laser as a pump and the
x~rays as a probe has received a lot of attention, \eg,
Ref.~\onlinecite{Wuilleumier:PP-06}.
For higher laser intensities, ionization of atoms takes place, producing
aligned, unoccupied atomic orbitals.
They are probed by using \textsc{xuv}~light or x~rays to excite an inner-shell
electron into them~\cite{Young:XR-06,Santra:SO-06,Loh:QS-07,Santra:SF-07}.
Second, there is a pump-probe setting where the laser pulse succeeds
the x~rays.
This situation has not received much of a focus
(see, however, Ref.~\onlinecite{Gagnon:SX-07}).
Third, there is a simultaneous exposition of atoms and molecules to two colors.
Recently, we studied molecules exposed to a laser with an intensity
close to but still below the excitation and ionization threshold.
If the molecule has an anisotropic polarizability tensor, they may be aligned
along the linear laser polarization axis~\cite{Santra:SF-07,Buth:LA-08,%
Peterson:XR-08,Buth:MD-08,Buth:US-up};
the x~rays serve as an \emph{in situ} probe of the rotational molecular
dynamics.
Going a bit higher in laser intensity, without exciting and ionizing, is
possible for rare gas atoms.
In krypton, laser dressing caused a pronounced deformation of the
photoabsorption cross section around the
$K$~edge~\cite{Buth:TX-07,Santra:SF-07}.
In neon we even found a strong suppression of the x-ray absorption
on the Ne$\,1s \to 3p$~resonance~\cite{Buth:ET-07,Santra:SF-07,Buth:RA-up,%
Buth:US-up}.

The name electromagnetically induced
transparency~(EIT)~\cite{Harris:NL-90,Boller:ET-91,Harris:EI-97}
for x~rays was coined to describe the strong reduction of absorption in neon.
It is related to EIT for optical wavelengths~\cite{Hau:SL-99,Phillips:SL-01,%
Lukin:ET-01,Fleischhauer:ET-05,Santra:HA-05} by being essentially given
by the same $\Lambda$-type three-level model.
The only difference in the case of EIT for x~rays is the linewidth of
the second unoccupied level which is not approximately zero.
Due to high laser intensity, transparency over a large number of wavelengths
arises.
The difference between the x-ray absorption spectra for laser-dressed neon
and krypton is predominantly due to the ten times higher decay width of
the $K$~vacancies in krypton ($\Gamma\I{1s} = 2.7
\eV$~\cite{Krause:NW-79,Chen:RK-80}) compared with neon ($\Gamma\I{1s} =
0.27 \eV$~\cite{Schmidt:ES-97}).
$K$~vacancies are produced by x-ray absorption and decay by x-ray fluorescence
and Auger decay~\cite{Thompson:XR-01}.
Another difference between EIT for optical wavelengths and EIT for x~rays lies
in the fact that refraction and dispersion by matter in the x-ray regime is
generally tiny.

EIT enables ultrafast control of the absorption and dispersion of a gaseous
medium on a femtosecond time scale because the coupling laser, typically
a Ti:sapphire laser system, can be combined with the sophisticated pulse
shaping technologies available for optical wavelengths~\cite{Weiner:FS-00}.
These pulse shapes can be imprinted on x~rays leading to
ultrafast pulse shaping technology for
x~rays~\cite{Buth:ET-07,Santra:SF-07,Buth:RA-up,Buth:US-up}.
The ability to create almost arbitrarily shaped x-ray pulses opens up
perspectives for the control of the dynamics of the inner-shell
electrons in atoms.
This way of shaping x-ray pulses is similar to
our recent study of x-ray absorption by laser-aligned molecules
where x-ray pulses are shaped by controlling the molecular
alignment~\cite{Buth:MD-08}.
As the rotational dynamics of molecules takes place on a picosecond
time scale, the x-ray pulse shaping is, however, done on a much slower
time scale.
Electron bunch manipulation techniques for short x-ray pulse generation are
presently developed at synchrotron radiation facilities~\cite{Borland:SA-05}.
Our approach complements these efforts in a very cost-effective way.

In this work, we would like to supplement our previous investigations
with a detailed account of the refraction and the dispersion of x~rays by
laser-dressed atoms.
Our theory is applied to argon.
Also we demonstrate that there is appreciable EIT for x~rays for argon.
As there is no substantial EIT effect for krypton, this means that EIT
for hard x~rays is predicted for the first time (the $K$~electrons of argon
can be ionized with x~rays of a wavelength
of~$3.9 \angstrom$~\cite{Thompson:XR-01}).
Such wavelengths are already useful to resolve coarse molecular structures
via x-ray diffraction in molecular imaging;
our work offers a way of temporal control of the x-ray pulses
in such experiments.

The paper is structured as follows.
In Sec.~\ref{sec:theory}, we evolve our quantum electrodynamic formalism of
Ref.~\onlinecite{Buth:TX-07} to incorporate rudimentary many-electron effects.
We determine the dynamic dipole polarizability for x~rays and the x-ray absorption
cross section in Secs.~\ref{sec:atelfield}, \ref{sec:dynstark}, and
\ref{sec:HFSapprox} on an \emph{ab initio} level.
Resolvents are evaluated in Sec.~\ref{sec:lanczos} with a single-vector
Lanczos algorithm.
Using these quantities together with the classical Maxwell equations in
Sec.~\ref{sec:refidx}, we devise the complex index of refraction for x~rays
of laser-dressed atoms.
The propagation of pulses through this new medium is treated in
Sec.~\ref{sec:pulseprop}.
Computational details are given in Sec.~\ref{sec:compdet} for
the results which are presented in Sec.~\ref{sec:results}.
Conclusions are drawn in Sec.~\ref{sec:conclusion}.
Finally, secondary physical processes of the pulse propagation from
Sec.~\ref{sec:pulseprop} are treated in the appendix.

Our equations are formulated in atomic units~\cite{Szabo:MQC-89}.
The Bohr radius~$1 \U{bohr} = 1 \, a_0$ is the unit of length
and $1 \, t_0$ represents the unit of time.
The unit of energy is~$1 \U{hartree} = 1 \Hartree$.
Intensities are given in units of~$1 \Hartree \> t_0^{-1} \, a_0^{-2}
= 6.43641 \times 10^{15} \U{W \, cm^{-2}}$.
Electric polarizabilities are measured in~$1 \, e^2 \, a_0^2 \, \Hartree^{-1}
= 1.64878 \E{-41} \U{C^2 \, m^2 \, J^{-1}}$.
The dielectric constant of the vacuum is~$\varepsilon_0 = \frac{1}{4\pi}$.

\section{Theory}
\label{sec:theory}
\subsection{Atoms in an electromagnetic field}
\label{sec:atelfield}

The time-independent Schr\"odinger equation of the field-free atom is
\begin{equation}
  \label{eq:Schroedi}
  \hat H\I{AT} \, \Psi^Z_s(\vec r_1 \, \sigma_1, \ldots, \vec r_Z \, \sigma_Z,
    t) = E_s \, \Psi^Z_s(\vec r_1 \, \sigma_1,
    \ldots, \vec r_Z \, \sigma_Z, t) \; .
\end{equation}
Here, $\hat H\I{AT}$~is the electronic Hamiltonian of the atom
which contains the Coulomb interaction of the $Z$~atomic
electrons with the nucleus and the two-particle interaction among the
electrons~\cite{Szabo:MQC-89,Merzbacher:QM-98}.
The wave functions~$\Psi^Z_s(\vec r_1 \, \sigma_1, \ldots, \vec r_Z \,
\sigma_Z, t)$ with eigenenergy~$E_s$ represent the ground state
($s = 0$) and core-excited excited states ($s \geq 1 $) of the
atom~\cite{footnote2} for the $u$th~electron at position~$\vec r_u$ with
spin projection quantum number~$\sigma_u$ for~$u \in \{1, \ldots, Z\}$.

The free electromagnetic field is represented by the Hamiltonian
\begin{subeqnarray}
  \label{eq:H_EM}
  \hat H_{\mathrm{EM}} &=& \Sum_{\vec k} \Sum_{\lambda = 1}^2
    \hat H_{\mathrm{EM}, \vec k, \lambda} \; , \\
  \hat H_{\mathrm{EM}, \vec k, \lambda} &=& \omega^{\vphantom{\dag}}_{\vec k} \,
    \hat a^{\dag}_{\vec k, \lambda} \hat a^{\vphantom{\dag}}_{\vec k, \lambda}
    - N^{\vphantom{\dag}}_{\vec k, \lambda} \,
    \omega^{\vphantom{\dag}}_{\vec k} \; ,
\end{subeqnarray}
where $\hat a^{\dag}_{\vec k, \lambda}$ and $\hat a^{\vphantom{\dag}}_{\vec k,
\lambda}$ denote creation and annihilation operators, respectively,
of photons in the mode~$\vec k, \lambda$ with wave vector~$\vec k$,
polarization index~$\lambda$, and energy~$\omega_{\vec k}$~\cite{Craig:MQ-84}.
Here, $N_{\vec k, \lambda}$ indicates the number of photons in the
mode~$\vec k, \lambda$ of some initial state.
The energy of the initial state, Eq.~(\ref{eq:Istate}) in our case,
has been adjusted such that the field energy is zero.
This simplifies our equations notably.
However, physical quantities are of course independent of the absolute
energy shift in Eq.~(\ref{eq:H_EM}).

We consider in this paper only two modes for the electromagnetic field
denoted by~L and X for the laser and the x~rays, respectively.
The eigenstates of the free electromagnetic field~$\hat H_{\mathrm{EM}}$
are Fock number states
\begin{equation}
  \ket{N\I{L}} \otimes \ket{N\I{X}} = (\hat a\I{L}^{\dag})^{N\I{L}} \,
    (\hat a\I{X}^{\dag})^{N\I{X}} \, \ket{0} \; ,
\end{equation}
with the vacuum state~$\ket{0}$.

The interaction of electrons with light is described by
\begin{equation}
  \label{eq:Hintlight}
  \hat H\I{I} = \Sum_{\sigma = -\frac{1}{2}}^{\frac{1}{2}} \ \Int_{\mathbb R^3}
    \hat \psi_{\sigma}^{\dagger}(\vec r) \; [\hat{\vec p} \mul \vec A(\vec r)
    + \dfrac{1}{2} \, \vec A^2(\vec r)] \; \hat \psi_{\sigma}(\vec r)
    \differential^3 r \; ,
\end{equation}
using the principle of minimal coupling to the electromagnetic
field~\cite{Craig:MQ-84,footnote1}.
The field is represented by the vector potential~$\vec A(\vec r)$
for which we assume the Coulomb gauge.
The electrons are created and annihilated by the field operators~$\hat
\psi_{\sigma}^{\dagger}(\vec r)$ and $\hat \psi_{\sigma}(\vec r)$,
respectively~\cite{Merzbacher:QM-98}.
We assume that both laser and x-ray wavelengths are sufficiently large for
the electric dipole approximation to be adequate.
Then, the vector potential becomes independent of~$\vec r$.

In dipole approximation, the form~(\ref{eq:Hintlight}) of the interaction
Hamiltonian with the x~rays reads
\begin{equation}
  \label{eq:Hintvelocity}
  \hat H\I{I,X} = \Sum_{\sigma = -\frac{1}{2}}^{\frac{1}{2}} \ \Int_{\mathbb R^3}
    \hat \psi_{\sigma}^{\dagger}(\vec r) \> [ \hat{\vec p} \mul \vec A\I{X}
    + \dfrac{1}{2}
    \vec A\I{X}^2 ] \> \hat \psi_{\sigma}(\vec r)
    \differential^3 r \; ,
\end{equation}
using the following mode expansion for the quantized vector potential
of the x~rays~\cite{Craig:MQ-84}:
\begin{equation}
  \label{eq:Amodeexp}
  \vec A\I{X} = \sqrt{\frac{2\pi}{V \, \omega\I{X}}} \; [ \vec e\I{X} \,
    \hat a\I{X} + \vec e\I{X}^{\,*} \, \hat a\I{X}^{\dagger}] \; .
\end{equation}
It is frequently referred to as velocity form~\cite{Scully:QO-97}.

In laser physics, one typically transforms the
interaction~(\ref{eq:Hintlight}) in dipole approximation to the so-called
length form.
This can, of course, be accomplished either for the Hamiltonian with
semiclassical electromagnetic fields~\cite{Scully:QO-97} or, as in our case,
in the quantum electrodynamic framework~\cite{Craig:MQ-84}.
Then, the interaction Hamiltonian with the laser field becomes
\begin{equation}
  \label{eq:Hintlength}
  \hat H\I{I,L} = \Sum_{\sigma = -\frac{1}{2}}^{\frac{1}{2}} \
    \Int_{\mathbb R^3} \hat \psi_{\sigma}^{\dagger}(\vec r)
    \> \vec r \mul \vec E\I{L} \> \hat \psi_{\sigma}(\vec r)
    \differential^3 r \; ,
\end{equation}
where we use the quantized mode expansion
\begin{equation}
  \label{eq:Emodeexp}
  \vec E\I{L} = \imag \, \sqrt{\frac{2\pi \, \omega\I{L}}{V}} \; [ \vec e\I{L}
    \, \hat a\I{L} - \vec e^{\,*}\I{L} \, \hat a\I{L}^{\dagger} ]
\end{equation}
for the electric field of the laser mode~\cite{Craig:MQ-84}.
The influence of the two-color light~(\ref{eq:Hintlight}) in dipole
approximation has been decomposed into~$\hat H\I{I} = \hat H\I{I,L}
+ \hat H\I{I,X}$ [Eqs.~(\ref{eq:Hintvelocity}) and (\ref{eq:Hintlength})].

The full Hamiltonian of an atom in two-color light~$\hat H = \hat H_0 + \hat H_1$
consists of a strongly interacting part~\cite{Buth:TX-07}
\begin{subeqnarray}
  \label{eq:H_0}
  \hat H_0      &=& \hat H\I{0,L} + \hat H\I{0,X} \; , \\
  \slabel{eq:H_0L}
  \hat H\I{0,L} &=& \hat H\I{AT} + \hat H\I{EM,L} + \hat H\I{I,L} \; , \\
  \hat H\I{0,X} &=& \hat H\I{EM,X} \; ,
\end{subeqnarray}
and a weak perturbation due to the interaction with the x~rays,~$\hat H_1 =
\hat H\I{I,X}$~\cite{Buth:TX-07}.

\subsection{Dynamic Stark effect}
\label{sec:dynstark}

We would like to determine the energy of the $K$-shell electrons in the
two-color field.
It is given by a perturbative expansion with respect to the x-ray field
using a basis of laser-dressed energy levels of the atom.
The initial state of the atom in light is given by the direct product of
the electronic ground state wave function and the Fock states
for the laser and the x-ray mode
\begin{subeqnarray}
  \label{eq:Istate}
  \ket{i} &=& \ket{\Psi^Z_0} \otimes \ket{N\I{L}} \; , \\
  \slabel{eq:IstateFull}
  \ket{I} &=& \ket{i}        \otimes \ket{N\I{X}} \; .
\end{subeqnarray}
The energy of the initial state is~$E_{I,0} = \bra{I} \hat H_0 \ket{I}$.
We assume that the electronic ground state of the atom is not noticeably
altered by the light.
Core-excited states with a vacancy in the $K$~shell serve as final states
for x-ray absorption and emission.
We define for~$s \neq 0$:
\begin{subeqnarray}
  \label{eq:FloBasis}
  \slabel{eq:FloBasisLas}
  \ket{\Phi_{s,\mu}} &=& \ket{\Psi_s^Z} \otimes \ket{N\I{L} - \mu} \; , \\
  \slabel{eq:FloBasisFull}
  \ket{\Phi^{\pm}_{s,\mu}} &=& \ket{\Phi_{s,\mu}} \otimes
    \ket{N\I{X} \pm 1} \; .
\end{subeqnarray}

Let us solve the strongly interacting laser-only problem~(\ref{eq:H_0L}) first
in terms of a non-Hermitian, complex-symmetric representation
of the Hamiltonian~\cite{Kukulin:TR-89,Moiseyev:CS-98,Santra:NH-02},
using the laser-only basis~(\ref{eq:FloBasisLas}) for~$s, s' \neq 0$:
\begin{equation}
  \label{eq:FloH0matrix}
  (\mat H\I{0,L})_{s,\mu ; s',\mu'} = \bra{\Phi_{s,\mu}} \hat H\I{0,L}
    \ket{\Phi_{s',\mu'}} \; .
\end{equation}
The eigenstates~$\ket{F}$ of Eq.~(\ref{eq:FloH0matrix}) form the
laser-dressed atomic energy levels with complex eigenenergies~$E_F$.

The interaction with the x-ray field is treated as a perturbation of the
laser-free atomic ground state;
the impact of the laser dressing arises exclusively in the final states.
The full basis states that include the x~rays are defined in
Eq.~(\ref{eq:FloBasisFull});
the laser-dressed atomic levels are given by the direct product
states~$\ket{F^{\pm}} = \ket{F} \otimes \ket{N\I{X} \pm 1}$
which correspond to the energies~$E_F^{\pm} = \bra{F^{\pm}} \hat H_0
\ket{F^{\pm}} = E_F \pm \omega\I{X}$.
We use non-Hermitian perturbation theory~\cite{Buth:NH-04,Buth:TX-07}
to determine the complex energy
\begin{equation}
  \label{eq:siegert}
  E\I{res} = E\I{R} - \imag \, \Gamma / 2
\end{equation}
of the initial state~(\ref{eq:IstateFull}), the so-called Siegert
energy~\cite{Kukulin:TR-89,Siegert:DF-39}.
The real part is the energy shift of the atomic level due to the laser
and x-ray fields~\cite{Sakurai:MQM-94,Buth:NH-04} whereas
$\Gamma$ stands for the transition rate from the initial state to
Rydberg states or the continuum.
The energy~(\ref{eq:siegert}) is obtained from
\begin{equation}
  \label{eq:srpt}
  \begin{array}{rcl}
    \displaystyle E_I &=& \displaystyle E_{I,0} + \underbrace{\bra{I} \hat H_1
      \ket{I}}_{P_1(\omega\I{X})} \\[5ex]
    &&\displaystyle{} + \underbrace{\sum_F \sum_{s
      \in \{+,-\}} \frac{\bra{I} \hat H_1 \ket{F^s} \bra{F^s} \hat H_1
      \ket{I}} {E_{I,0} - E_F^s}}_{P_2(\omega\I{X})} \; .
  \end{array}
\end{equation}
Assuming~$N\I{X} \gg 1$, the interaction matrix elements~$\bra{I}
\hat H_1 \ket{F^+}$ and $\bra{I} \hat H_1 \ket{F^-}$ are approximately
the same.

The first order correction~$P_1(\omega\I{X})$ in Eq.~(\ref{eq:srpt}) is found by
combining Eqs.~(\ref{eq:Hintvelocity}), (\ref{eq:Amodeexp}), and
(\ref{eq:Istate}):
\begin{equation}
  \label{eq:P_1_shift}
  \begin{array}{rcl}
  \displaystyle P_1(\omega\I{X}) &=& \displaystyle \frac{1}{2} \, \bra{I}
    \Sum_{\sigma = -\frac{1}{2}}^{\frac{1}{2}} \ \Int_{\mathbb R^3}
    \hat \psi_{\sigma}^{\dagger}(\vec r) \> \vec A^2\I{X} \> \hat
    \psi_{\sigma}(\vec r) \differential^3 r \ket{I} \\
    &=& \displaystyle \frac{\pi \, Z}{V \, \omega\I{X}} \, (2 \, N\I{X} + 1)
    \approx \frac{2\pi}{V \, \omega\I{X}} \, N\I{X} \, Z \; .
  \end{array}
\end{equation}
The one-particle operator~$\vec A^2\I{X}$ hereby acts on all $Z$~electrons
in the initial state~(\ref{eq:IstateFull}).
The tiny correction~$\frac{\pi \, Z}{V \, \omega\I{X}}$ after the second equals
sign vanishes upon taking the limit~$V \to \infty$ in the end.
The real quantity~$P_1(\omega\I{X})$ in Eq.~(\ref{eq:P_1_shift}) is the
energy shift due to the scattering of x~rays and represents the ac~Stark
shift~\cite{Delone:AC-99,Buth:NH-04}.
We find the corresponding dynamic polarizability~\cite{Delone:AC-99} from
\begin{equation}
   \label{eq:P_1_polar}
   \re P_1(\omega\I{X}) = -\frac{1}{4} \, \alpha_1(\omega\I{X})
     \, E\I{X,0}^2 \; .
\end{equation}
The peak electric field of the x~rays is~$E\I{X,0} = \sqrt{8\pi \, \alpha \,
I\I{X}}$~\cite{Scully:QO-97,Meystre:QO-91} for the x-ray intensity~$I\I{X}$.
From Eqs.~(\ref{eq:P_1_shift}) and (\ref{eq:P_1_polar}), we obtain the first
order atomic polarizability
\begin{equation}
  \label{eq:polarfirst}
  \alpha_1(\omega\I{X}) = - {Z \over \omega\I{X}^2} \; .
\end{equation}

In analogy to Eq.~(\ref{eq:P_1_polar}), we define the second order
contribution to the dynamic polarizability~\cite{Delone:AC-99}.
With the real part of~$P_2(\omega\I{X})$ in Eq.~(\ref{eq:srpt}),
the well-known Kramers-Heisenberg form
\begin{equation}
  \label{eq:Kramers-Heisenberg}
  \begin{array}{rcl}
    \displaystyle \alpha_2(\omega\I{X}) &=& \displaystyle -\dfrac{4}
      {E\I{X,0}^2} \  \re P_2(\omega\I{X}) \\
      &=& \displaystyle \dfrac{4}{E\I{X,0}^2} \  \re \Sum_F
      \bra{I} \hat H_1 \ket{F^+} \, \bra{F^+} \hat H_1 \ket{I} \\
      &&\displaystyle \hspace{5em} {} \times\,
      \dfrac{2 \, (E_F - E_{I,0})}{(E_F - E_{I,0})^2 - \omega^2\I{X}}
  \end{array}
\end{equation}
results~\cite{Craig:MQ-84,Delone:AC-99}.

Using the first and second order atomic polarizabilities from
Eqs.~(\ref{eq:polarfirst}) and (\ref{eq:Kramers-Heisenberg}),
we can define the total atomic polarizability
\begin{equation}
  \label{eq:polartot}
  \alpha\I{I}(\omega\I{X}) = \alpha_1(\omega\I{X}) + \alpha_2(\omega\I{X}) \; .
\end{equation}
We derive explicit equations for the polarizability~$\alpha_2(\omega\I{X})$
in the next section.

\subsection{Hartree-Fock-Slater approximation}
\label{sec:HFSapprox}

We solve the Schr\"odinger equation~(\ref{eq:Schroedi}) with the help of the
independent-electron approximation~\cite{Szabo:MQC-89,Merzbacher:QM-98,%
Buth:TX-07}.
Then, the $Z$-electron ground-state wave function is given by a Slater
determinant of one-electron orbitals~$\Phi^Z_0(\vec r_1 \, \sigma_1,
\ldots, \vec r_Z \, \sigma_Z)$.
Using the formalism of second quantization, we have
\begin{equation}
  \label{eq:SlaterDet}
  \ket{\Phi^Z_0} = \prod_{n,l,m \  \mathrm{occupied}} \hat
    b^{\dagger}_{n,l,m,\uparrow} \, \hat b^{\dagger}_{n,l,m,\downarrow}
    \ket{0}
\end{equation}
for a closed-shell atom.
Here, $\hat b^{\dagger}_{n,l,m,\uparrow}$ and
$\hat b^{\dagger}_{n,l,m,\downarrow}$ are creators of electrons
in the spatial orbital~$\varphi_{n,l,m}(\vec r)$ with spin projection
quantum number up ($\sigma = \frac{1}{2} = {\uparrow}$) and down
($\sigma =-\frac{1}{2} = {\downarrow}$), respectively.
The spatial orbitals are eigenfunctions of the one-electron Hamiltonian
\begin{equation}
  \hat H\I{HFS} = -\frac{1}{2} \, \vec \nabla^2 + V\I{HFS}(r) \; .
\end{equation}
They are characterized by the principal quantum number~$n$, the angular
momentum~$l$, and the magnetic quantum number~$m$~\cite{Merzbacher:QM-98}.
To determine the effective one-electron central potential~$V\I{HFS}(r)$,
we made the Hartree-Fock-Slater mean-field
approximation~\cite{Slater:AS-51,Slater:XA-72}.

When x~rays are absorbed, an electron may be ejected into the continuum
leaving a core hole behind~\cite{footnote3}.
We use a complex absorbing potential~(CAP) to handle such continuum
electrons.
The CAP is a one-particle operator
which is added to~$\hat H\I{HFS}$.
It is derived from smooth exterior complex scaling~\cite{Moiseyev:DU-98,%
Riss:TC-98,Karlsson:AR-98,Buth:TX-07}.
Additionally, one needs to allow for the relaxation of core holes by x-ray
fluorescence and Auger decay~\cite{Thompson:XR-01,Als-Nielsen:EM-01} with
a decay width of~$\Gamma\I{1s}$.
According to Eq.~(\ref{eq:siegert}), this is accounted for by
adding~$-\imag \, \Gamma\I{1s} / 2$ to all energies
of core-excited states~\cite{Buth:TX-07}.
Subsuming these contributions, we obtain the effective one-electron
atomic Hamiltonian~\cite{Buth:TX-07} which we use instead of~$\hat H\I{AT}$
in Eq.~(\ref{eq:H_0L}).

With the help of the spin orbitals, we can expand the field
operators~\cite{Merzbacher:QM-98} in the equations of
Secs.~\ref{sec:atelfield} and \ref{sec:dynstark} as follows
\begin{subeqnarray}
  \hat \psi_{\sigma}(\vec r) &=& \Sum_{n,l,m} \varphi_{n,l,m}(\vec r)
    \> \hat b_{n,l,m,\sigma} \; , \\
  \hat \psi_{\sigma}^{\dagger}(\vec r) &=& \Sum_{n,l,m}
    \varphi^*_{n,l,m}(\vec r) \> \hat b^{\dagger}_{n,l,m,\sigma} \; .
\end{subeqnarray}
Based on the independent-particle ground state~(\ref{eq:SlaterDet}), we
construct the spin-singlet $K$-shell-excited
states~\cite{Szabo:MQC-89,Merzbacher:QM-98}:
\begin{subeqnarray}
  \ket{\Phi_{n,l,m,\mu}} &=& \dfrac{1}{\sqrt{2}} \> [\hat b^{\dagger}_{n,l,m,
    \uparrow} \> \hat b^{\vphantom{\dagger}}_{1,0,0,\uparrow} \nonumber \\
  &&\hspace{2em} {} + \hat b^{\dagger}_{n,l,m,\downarrow} \> \hat b^{\vphantom
    {\dagger}}_{1,0,0,\downarrow}] \ket{\Phi^Z_0} \\
  &&{} \otimes \ket{N\I{L} - \mu} \; , \nonumber \\
  \slabel{eq:FloBasisFullHFS}
  \ket{\Phi^{\pm}_{n,l,m,\mu}} &=& \ket{\Phi_{n,l,m,\mu}} \otimes
    \ket{N\I{X} \pm 1} \; ,
\end{subeqnarray}
as basis states [see Eq.~(\ref{eq:FloBasis})].
Note that $m$ is a conserved quantum number because the laser is
linearly polarized;
thus the matrix~(\ref{eq:FloH0matrix}) blocks with respect to~$m$.
Following Ref.~\onlinecite{Buth:TX-07}, the Kramers-Heisenberg
form~(\ref{eq:Kramers-Heisenberg}) becomes
\begin{equation}
  \label{eq:polarizability}
  \begin{array}{rcl}
    \displaystyle \alpha_2(\omega\I{X}, \vartheta\I{LX}) &=& \displaystyle
      \frac{2}{3} \ \re \biggl[ \Sum_{m=-1}^1 \varkappa_m (\vartheta\I{LX}) \\
    &&\displaystyle \hspace{2.2em} {} \times \Sum_F {(d_F^{(m)})^2 \over
      E_F^{(m)} - E_{1s} - \omega\I{X}} \biggr] \; .
  \end{array}
\end{equation}
Here, we omit the emission term~``$+$'' which results from decomposing the
denominator of Eq.~(\ref{eq:Kramers-Heisenberg}) to emphasize the
similarity to the expression for the cross section.
The initial-state energy is the $K$-shell energy~$E_{I,0} = E_{1s}$.
The angular dependence of the polarizability~(\ref{eq:polarizability}) is given
by~$\varkappa_m(\vartheta\I{LX})$ which is equal to~$\cos^2 \vartheta\I{LX}$
for~$m = 0$ and equal to $\frac{1}{2} \sin^2 \vartheta\I{LX}$ for~$m = \pm 1$.
The radial dipole matrix elements are given by~$d_F^{(m)}$~\cite{Buth:TX-07}.

\subsection{Lanczos solution of the polarizability}
\label{sec:lanczos}

The full diagonalization of the interaction of the atom with the laser
to obtain the new energy levels of a laser-dressed atom becomes
impractical for larger numbers of atomic orbitals and photons.
To be able to solve the equations for the
polarizability~(\ref{eq:polarizability}) and the cross
section~\cite{Buth:TX-07} with acceptable computational effort, we employ
a single-vector Lanczos algorithm~\cite{Lanczos:IM-50,Cullum:LA-85}.
A real symmetric version has been implemented by Meyer and
Pal~\cite{Meyer:BL-89};
it was extended to complex symmetric matrices by
Sommerfeld~\etal~\cite{Sommerfeld:TA-98}.

The expression for~$P_2(\omega\I{X})$ in Eq.~(\ref{eq:srpt}) can be
rewritten as
\begin{equation}
  \label{eq:P_2_op}
  P_2(\omega\I{X}) = \bra{I} \hat H_1 \hat P \; \frac{1}{E_{1s} - \hat H_0} \;
    \hat P H_1 \ket{I} \; ,
\end{equation}
with the projector on the core-excited basis states~(\ref{eq:FloBasisFullHFS}):
\begin{equation}
  \label{eq:G_I_td}
  \hat P = \hat P^+ + \hat P^- = \Sum_{\scriptstyle n, l, m, \mu, \atop
    \scriptstyle s \in \{+,-\}} \ket{\Phi^s_{n,l,m,\mu}}
    \bra{\Phi^s_{n,l,m,\mu}} \; .
\end{equation}
Let us define the resolvents
\begin{equation}
  \label{eq:resolvent}
    \displaystyle R^{(m),\pm}_{n,l,\mu ; n',l',\mu'} = \displaystyle
      \bra{\Phi^{\pm}_{n,l,m,\mu}} \frac{1}{E_{1s} - \hat H_0}
      \; \ket{\Phi^{\pm}_{n',l',m,\mu'}} \; ;
\end{equation}
the two cases for x-ray absorption and emission are only distinguished by
a~$\mp \omega\I{X}$ term in the denominator.
Apart from this, the expressions involves only the interaction with the
laser.
In fact, the resolvents~(\ref{eq:resolvent}) can be approximated in terms
of a single-vector Lanczos run for the matrix~$\mat H\I{0,L}^{(m)}$
[Eq.~(\ref{eq:FloH0matrix})].
To show this, let the components of the start
vector~\cite{Cullum:LA-85,Meyer:BL-89} be
\begin{equation}
 \label{eq:lanczstart}
  v^{(m)}_{n,l,\mu} = \bra{\Phi^+_{n,l,m,\mu}} \hat H\I{I,X} \ket{I} \; .
\end{equation}
In Floquet approximation, the results are the same for the two cases~``$+$''
and ``$-$'' of x-ray emission and absorption;
we use the former case throughout.
Using Eqs.~(\ref{eq:G_I_td}), (\ref{eq:resolvent}), and (\ref{eq:lanczstart}),
expression~(\ref{eq:P_2_op}), is recast into
\begin{equation}
  \label{eq:P_2_resol}
  P_2(\omega\I{X}) = \Sum_m \vec v^{(m)}\transpose \> [\mat R^{(m),+} +
    \mat R^{(m),-}] \, \vec v^{(m)} \; .
\end{equation}

After $N\I{iter}$ Lanczos iterations, we have reduced~$\mat H\I{0,L}^{(m)}$
to the tridiagonal matrix~$\mat T^{(m)} = \mat Q^{(m)}\transpose \mat
H\I{0,L}^{(m)} \mat Q^{(m)}$ where $\mat Q^{(m)}$~is the matrix of Lanczos vectors.
The matrix~$\mat T^{(m)}$ has the eigenvalues~$\mat \Lambda^{(m)}$, and the
eigenvectors~$\mat X^{(m)}$, \ie, we have~$\mat X^{(m)}\transpose \mat T^{(m)}
\mat X^{(m)} = \mat \Lambda^{(m)}$.
We insert~$(\mat Q^{(m)}\mat X^{(m)})(\mat Q^{(m)}\mat X^{(m)})\transpose =
\unitmatrix$ on the left hand side of~$\vec v^{(m)}$ in
Eq.~(\ref{eq:P_2_resol}) and its transpose on the right hand side
of~$\vec v^{(m)}\transpose$.
The start vector~(\ref{eq:lanczstart}) is normalized at the beginning of the
Lanczos iterations.
Hence that~$\mat Q^{(m)}\transpose \, \vec v^{(m)} = \|\vec v^{(m)}\| \>
\vec e_1$, with~$\vec e_1 = (1, 0, \ldots, 0)\transpose$.
Finally, Eq.~(\ref{eq:P_2_resol}) reads in the Lanczos basis
\begin{equation}
  \label{eq:P_2_lancz}
  \begin{array}{rcl}
    \displaystyle P_2(\omega\I{X}) &=& \displaystyle \sum_m \vec v^{(m)}
      \transpose \, \vec v^{(m)} \  \sum_{p=1}^{N\I{iter}} (X^{(m)}_{1p})^2 \\
    &&\displaystyle{} \times \biggl[
      \frac{1}{E_{1s} + \omega\I{X} - \Lambda^{(m)}_{pp}} +
      \frac{1}{E_{1s} - \omega\I{X} - \Lambda^{(m)}_{pp}} \biggr] \; .
  \end{array}
\end{equation}
From~$P_2(\omega\I{X})$ in Eq.~(\ref{eq:P_2_lancz}), we obtain the
dynamic atomic polarizability~(\ref{eq:Kramers-Heisenberg}) by
taking the real part of~$P_2(\omega\I{X})$.
The photoabsorption cross section follows from the imaginary part
of~$P_2(\omega\I{X})$ using the prefactors in Ref.~\onlinecite{Buth:TX-07}.

\subsection{Index of refraction}
\label{sec:refidx}

We characterize the impact of a gaseous medium of laser-dressed
atoms on x~rays in terms of the complex index of refraction.
From the Maxwell equations, we obtain the equation for the electric
field~\cite{Scully:QO-97,Jackson:CE-98} of a wave propagating in the
$z$~direction:
\begin{equation}
  \label{eq:MaxWave}
  \frac{\partial^2 E(z,t)}{\partial z^2} - \frac{1}{c^2} \, \frac{\partial^2 E(z,t)}
    {\partial t^2} = \frac{4\pi}{c^2} \, \frac{\partial^2 P(z,t)}{\partial t^2} \; .
\end{equation}
Let $E(z,t) = E^+(z,t) + E^-(z,t)$ and $E^-(z,t) =
[E^+(z,t)]^*$~\cite{Scully:QO-97,Meystre:QO-91} with
\begin{equation}
  \label{eq:elecfwav}
  E^+(z,t) = \frac{1}{2} \, E_0 \, \euler^{\imag \, (kz - \omega\I{X} t)} \; .
\end{equation}
Here, $k$~is the wave number in the medium.
The complex polarization~$P(z,t)$ describes the linear response of an
atom to an electric field in terms of the complex electric
susceptibility~$\chi(\omega\I{X})$ via
\begin{equation}
  \label{eq:polmedium}
  P^{\pm}(z,t) = \frac{1}{4\pi} \, \chi(\omega\I{X}) \, E^{\pm}(z,t) \; .
\end{equation}
We define~$P(z,t) = P^+(z,t) + P^-(z,t)$, $P^-(z,t) = [P^+(z,t)]^*$,
and $P_0 = \frac{1}{4\pi} \, \chi(\omega\I{X}) \, E_0$.

Inserting~$P(z,t)$ and $E(z,t)$ into Eq.~(\ref{eq:MaxWave}) yields the
relation between the refractive index and the
polarization~\cite{Jackson:CE-98}:
\begin{equation}
  \label{eq:refindex}
  n^2(\omega\I{X}) \equiv 1 + \chi(\omega\I{X})
    \equiv \frac{k^2 c^2}{\omega\I{X}^2}
    = 1 + \frac{4\pi \, P_0}{E_0} \; .
\end{equation}
Consequently, the propagation through the medium is described by
Eq.~(\ref{eq:MaxWave}) in terms of
\begin{equation}
  \label{eq:elecfwavidx}
  E^+(z,t) = \frac{1}{2} \, E_0 \, \euler^{\imag \, [n(\omega\I{X}) \, k_0 z
    - \omega\I{X} t]} \; ,
\end{equation}
with the wave number in vacuum~$k_0 = \frac{\omega\I{X}}{c}$.
Equation~(\ref{eq:elecfwavidx}) is used to form a solution~$E(z,t) =
E^+(z,t) + [E^+(z,t)]^*$ of the wave
equation~(\ref{eq:MaxWave}), \ie, Eq.~(\ref{eq:refindex}) is satisfied
upon letting~$k = n(\omega\I{X}) \, k_0$.

To make the connection of our classical electrodynamic equations to our
quantum electrodynamic results from Sec.~\ref{sec:dynstark},
we use the dynamic polarizability~(\ref{eq:polartot}) to express
the polarization~(\ref{eq:polmedium}) of the medium with number
density~$n_{\#}$ due to the x~rays,
\begin{equation}
  \label{eq:polatpol}
  P_0 = n_{\#} \, \alpha\I{I}(\omega\I{X}) \, E_0
           + \frac{\imag \, \mu}{4 \pi \, k_0} \, E_0 \; ,
\end{equation}
where we add an imaginary intensity absorption term.
It contains the absorption coefficient~$\mu = n_{\#} \, \sigma(\omega\I{X})$
which involves the x-ray absorption cross section of
the atom~$\sigma(\omega\I{X})$~\cite{Als-Nielsen:EM-01}.
Using Eqs.~(\ref{eq:refindex}) and (\ref{eq:polatpol}), the index of
refraction becomes
\begin{equation}
  \label{eq:refractive}
  n(\omega\I{X}) \approx 1 + \frac{\chi(\omega\I{X})}{2} = 1 + 2\pi \,
    n_{\#} \, \alpha\I{I}(\omega\I{X}) + \imag \, \frac{\mu}{2\,k_0} \; ,
\end{equation}
where the square root of the right hand side of Eq.~(\ref{eq:refindex})
was approximated.
The imaginary part in Eqs.~(\ref{eq:polatpol}) and (\ref{eq:refractive})
leads together with Eq.~(\ref{eq:elecfwavidx}) to an exponential decay
following Beer's law~\cite{Meystre:QO-91,Als-Nielsen:EM-01}.

\subsection{Pulse propagation}
\label{sec:pulseprop}

We assume copropagating laser and x-ray pulses which pass through
a gas cell of length~$L$ that is filled with a gas of pressure~$p$
and temperature~$T$ (see Fig.~\ref{fig:ppset}).
To describe the evolution of the pulses, we employ the rate-equation
approximation~\cite{Meystre:QO-91}.
The rate equations are solved on a numerical grid of~$N\I{grd}$ points
to discretize the distance covered during the propagation time interval.
Let the laser radiation of angular frequency~$\omega\I{L}$ be linearly
polarized along the $z$~axis.
The laser intensity is given by~$I\I{L}(t)$.
It is assumed to be constant perpendicular to the beam axis.
We assume an x-ray pulse with a Gaussian envelope with peak flux~$J\I{X,0}$
and a full width at half maximum~(FWHM) duration of~$\tau\I{X}$:
\begin{equation}
  \label{eq:XrayGaussEnv}
  J\I{X}(t) = J\I{X,0} \> \euler^{-4 \ln 2 \, (\frac{t}{\tau\I{X}})^2} \; .
\end{equation}
The peak flux of the pulse (at~$t = 0$) depends on the number of photons per x-ray
bunch~$n\I{ph}$ like
\begin{equation}
  \label{eq:XrayPeakFlux}
  J\I{X,0} = 2 \, \sqrt{\frac{\ln 2}{\pi}} \> \frac{n\I{ph}}{\tau\I{X}}
    \, X(0) \; .
\end{equation}
The factor~$X(0) = \frac{4\ln 2}{\pi \varrho\I{X}^2}$ represents the peak of a
Gaussian radial profile of a FWHM width of~$\varrho\I{X}$.

\begin{figure}
  \begin{center}
    \includegraphics[clip,width=\hsize]{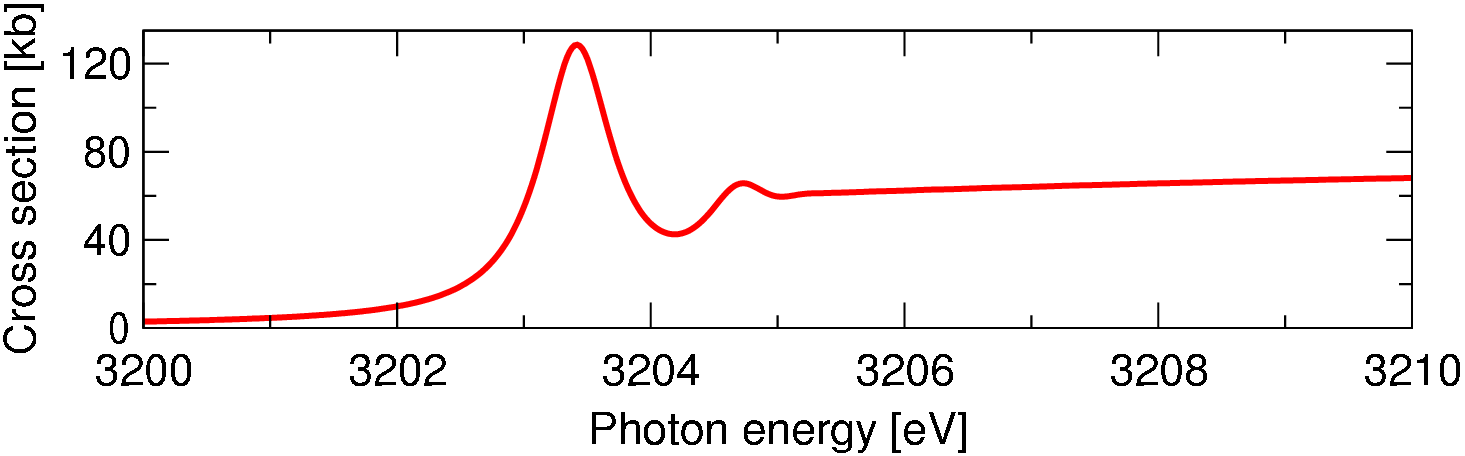}
    \caption{(Color online) X-ray absorption cross section of argon near
             the $K$~edge without laser dressing.}
    \label{fig:x-ray-abs}
  \end{center}
\end{figure}

We assume that the laser light and the x~rays both propagate with the speed
of light.
In the appendix, we made a number of estimates of the impact of
secondary physical effects in the gas cell on the phase and group velocities
of the laser and the x~rays.

We describe the interaction of the two-color light with the gas by the
following rate equation for the x-ray pulse
\begin{equation}
  \label{eq:rateflux}
  \Delta J\I{X}(z_{j+1}) = -n'_{\#}(z_j) \, \sigma'(z_j) \, J'\I{X}(z_j)
    \, \Delta z \; ,
\end{equation}
which is evaluated for the individual steps~$j \in \{1, \ldots, N\I{grd}-1\}$
to move the pulse over the grid.
Let $\sigma'(z_j) \equiv \sigma(I\I{L}(z_j / c))$ and $J'\I{X}(z_j)
\equiv J\I{X}(z_j / c)$.
Further, $n'_{\#}(z_j)$ represents the number density of argon atoms
at~$z_j$.
Initially, it is $n_{\#}$ inside the gas cell and zero otherwise.
Equation~(\ref{eq:rateflux}) can be solved analytically in the gas cell;
one obtains an exponential decay following Beer's
law~\cite{Meystre:QO-91,Als-Nielsen:EM-01}
for each infinitesimal section of constant intensity of the x-ray pulse.

\section{Computational details}
\label{sec:compdet}

The computations presented in the ensuing Sec.~\ref{sec:results} were carried
out using the \textsc{dreyd} and the \textsc{pulseprop} programs of the
\textsc{fella} package~\cite{fella:pgm-V1.3.0}.
We developed \textsc{dreyd} and \textsc{pulseprop} for
Refs.~\onlinecite{Buth:ET-07,Santra:SF-07,Buth:RA-up,Buth:US-up}.
For this paper, we added the treatment of the velocity
form~(\ref{eq:Hintvelocity}) of the interaction Hamiltonian with the
x~rays to \textsc{dreyd}.

The computational parameters of \textsc{dreyd} are specified following
Ref.~\onlinecite{Buth:TX-07}.
To solve the atomic electronic structure problem, we use the
Hartree-Fock-Slater code of Herman and Skillman~\cite{Herman:AS-63}
setting the $X\alpha$~parameter to unity.
The radial part of the atomic orbitals is represented on a grid of a radius
of~$60 \bohr$ using 3001~finite-element functions,
considering angular momenta up to~$l=7$.
We represent the radial Schr\"odinger equation in this basis set.
From its eigenfunctions, we choose, for each~$l$, the 100~lowest in energy
to form atomic orbitals~\cite{Buth:TX-07}.
The smooth exterior complex scaling complex absorbing potential
is formed using the complex scaling angle~$\theta  = 0.13 \rad$,
a smoothness of the path of~$\lambda = 5 \bohr^{-1}$, and a start
distance of~$r_0 = 7 \bohr$.
We think of a Ti:sapphire laser system as a potential dressing laser
which emits light at a wavelength of~$800 \U{nm}$, \ie, a photon energy
of~$\omega\I{L} = 1.55 \eV$.
To converge the x-ray absorption spectra, we accounted for the emission and
absorption of up to 12~laser photons
in the Floquet-type matrix~(\ref{eq:FloH0matrix}).
We set the $K$~edge of argon to its experimental value of~$E\I{1s}
= 3205.9 \eV$~\cite{Thompson:XR-01} as well as the linewidth of a
$K$~vacancy~$\Gamma\I{1s} = 0.66 \eV$~\cite{Campbell:WA-01}.
Finally, we need to carry out~$N\I{iter} = 4000$ Lanczos iterations
to converge~$P_2(\omega\I{X})$ in Eq.~(\ref{eq:P_2_lancz}).

\begin{figure}
  \includegraphics[clip,width=\hsize]{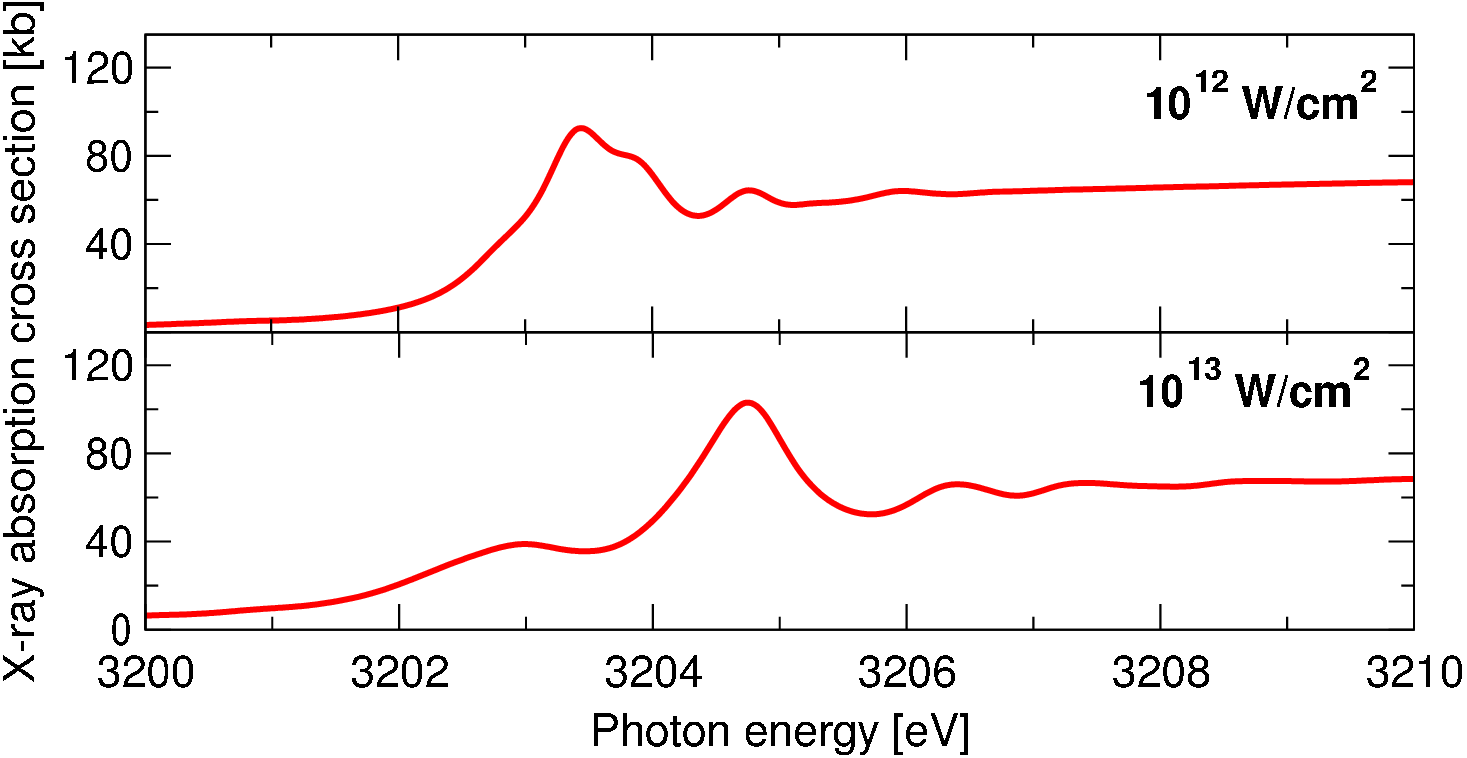}
  \caption{(Color online) X-ray absorption cross section of argon for parallel
           laser and x-ray polarization vectors.
           In the upper panel, the intensity of the dressing laser
           is~$10^{12} \U{W/cm^2}$;
           it is $10^{13} \U{W/cm^2}$ in the lower panel.}
  \label{fig:XLpar}
\end{figure}

The \textsc{pulseprop} code assumes a gas cell with a length of~$6 \U{mm}$
(Fig.~\ref{fig:ppset}).
It is filled with argon gas of a pressure of~$p = 5 \atm$ at a
temperature of~$T = 300 \U{K}$.
The number density of argon atoms follows from the ideal gas
law~$n_{\#} = \frac{p}{T}
= 1.22 \E{20} \U{cm}^{-3}$.
The x~rays are tuned to the Ar$\, 1s \to 4p$~resonance which has in our
computations an energy of~$3203.42 \eV$ (peak position in
Fig.~\ref{fig:x-ray-abs}).
Each x-ray pulse~(\ref{eq:XrayGaussEnv}) comprises $n\I{ph} = 10^6$~photons
and has a FWHM duration of~$\tau\I{X} = 100 \U{ps}$.
The x-ray beam is circular with a FWHM focal width of~$10^{-3} \U{cm}$.
This implies a peak x-ray flux~(\ref{eq:XrayPeakFlux}) of~$J\I{X,0} = 8.3 \E{9}
\U{ps^{-1} cm^{-2}}$.
The laser pulse envelope has the shape
\begin{equation}
  \label{eq:lasshape}
  \begin{array}{rcl}
    \displaystyle I\I{L}(z_0,z\I{L},z) &=& \displaystyle I\I{L,0} \,
      \Sum_{m=0}^4 \, \cos \Bigl( \frac{\pi}{16} \, m \Bigr) \\[4ex]
     &&\displaystyle{} \times \euler^{-4 \ln 2 \> [(z-z_0 + 2 \, (m-2)
       \, z\I{L}) / z\I{L}]^2} \; .
  \end{array}
\end{equation}
The FWHM length of the constituting five individual laser pulses is
defined by~$z\I{L} = c \, \tau\I{L}$ for the FWHM pulse
duration~$\tau\I{L} = 1 \U{ps}$.
The pulse train is centered at~$z_0$.

\section{Results and discussion}
\label{sec:results}

\begin{figure}
  \includegraphics[clip,width=\hsize]{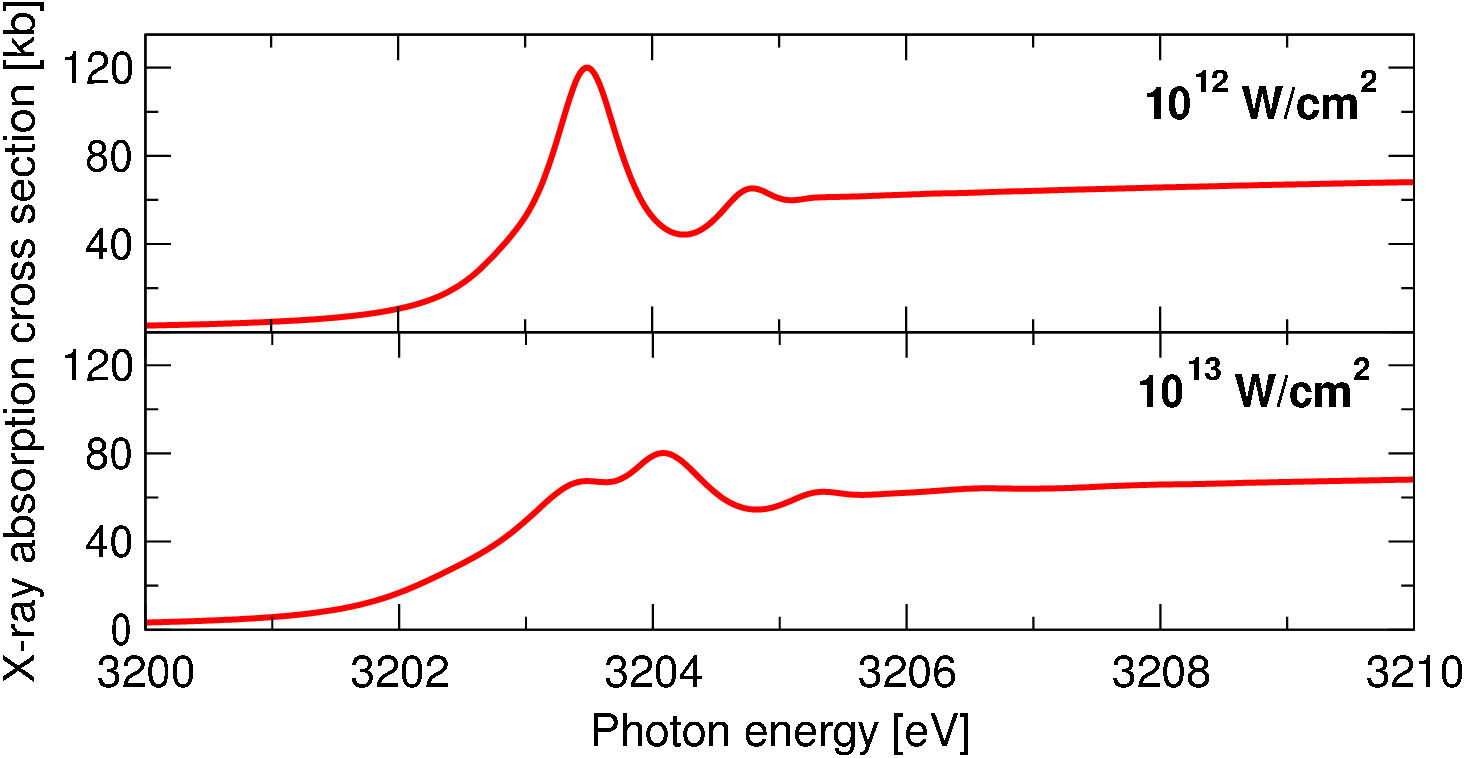}
  \caption{(Color online) X-ray absorption cross section of argon for
           perpendicular laser and x-ray polarization vectors.
           The dressing laser has an intensity of~$10^{12} \U{W/cm^2}$ in the
           upper panel and of~$10^{13} \U{W/cm^2}$ in the lower panel.}
  \label{fig:XLperp}
\end{figure}

In Fig.~\ref{fig:x-ray-abs}, we show the x-ray absorption cross section of
argon near the $K$~edge without laser dressing.
Clearly, the isolated Ar$\, 1s \to 4p$~pre-edge resonance is discernible
at~$3203.42 \eV$.
%
%
The much weaker Ar$\, 1s \to 5p$~pre-edge resonance can also be seen
at~$3204.72 \eV$.
The continuum stretches beyond the $K$~edge of argon which lies
at~$3205.9 \eV$~\cite{Thompson:XR-01}.

We expose the argon atoms to a linearly polarized
$800 \U{nm}$~laser with intensities up to~$10^{13} \U{W/cm^2}$.
The intensity remains well below the appearance intensity
of~Ar$^+$~\cite{Augst:TI-89}.
This is an important assertion because our theoretical description does not
account for ionization without prior x-ray absorption.

In Fig.~\ref{fig:XLpar}, we display the x-ray absorption cross section with
laser-dressing and parallel laser and x-ray polarization vectors.
The cross section is shown for the two laser intensities~$10^{12} \U{W/cm^2}$
and $10^{13} \U{W/cm^2}$.
The spectrum for~$10^{12} \U{W/cm^2}$ in the upper panel of Fig.~\ref{fig:XLpar}
resembles the one without laser in Fig.~\ref{fig:x-ray-abs}.
Yet the Ar$\, 1s \to 4p$~resonance is split into two close-lying peaks.
Also small new bumps emerge in the spectrum, \eg, around~$3206 \eV$.
For $10^{13} \U{W/cm^2}$, the trends become more pronounced;
the absorption spectrum is significantly modified compared to the laser-free
case.
The Ar$\, 1s \to 4p$~resonance is split into two peaks which are separated
by almost~$2 \eV$.

\begin{figure}
  \includegraphics[clip,width=5cm]{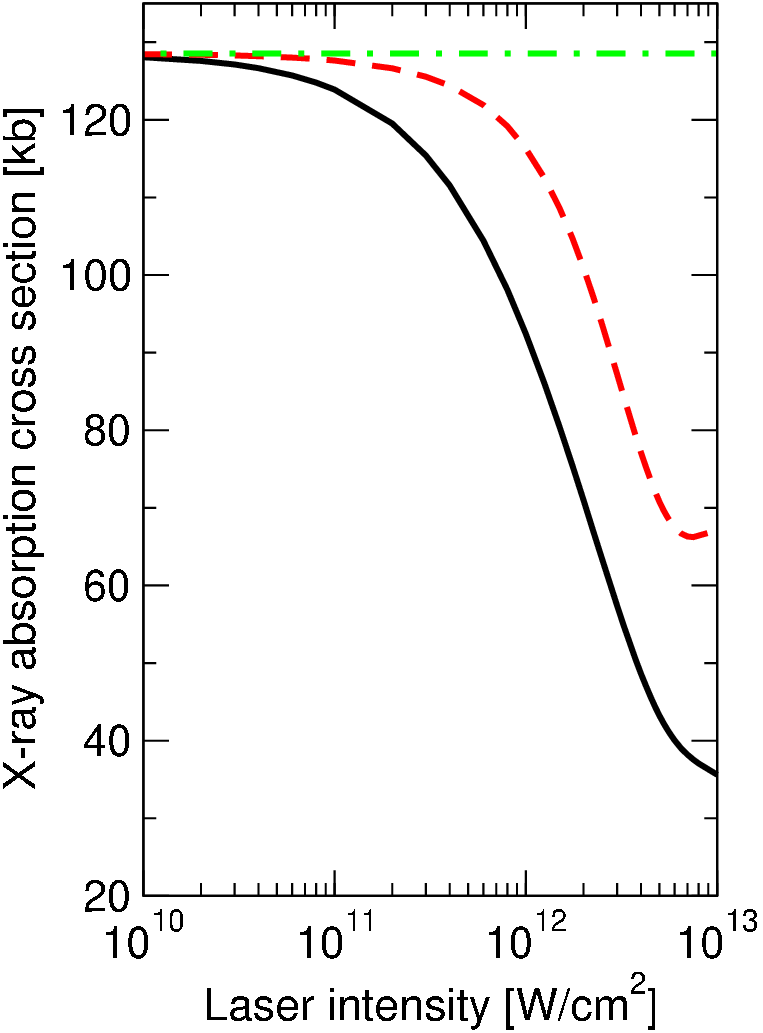}
  \caption{(Color online) Dependence on the intensity of the dressing laser of
           the x-ray absorption cross section of argon on the
           Ar$\,1s \to 4p$~pre-edge resonance at~$3203.42 \eV$ for parallel
           laser and x-ray polarization vectors (solid black curve) and
           for perpendicular vectors (dashed red curve).
           The dash-dotted green line represents the cross section without
           laser dressing.}
  \label{fig:sigma_argon}
\end{figure}

When one has tuned the x-ray energy to the peak of the Ar$\, 1s \to
4p$~resonance without laser dressing, then, turning on the laser, leads
to a substantial suppression of x-ray absorption.
This phenomenon was analyzed by us in detail in Ref.~\onlinecite{Buth:ET-07}.
We termed it electromagnetically induced transparency~(EIT) for x~rays.
The mechanism which leads to EIT for x~rays was analyzed in
Ref.~\onlinecite{Buth:ET-07} in terms of a quantum-optical three-level
model~\cite{Boller:ET-91}.
The model reproduces well our \emph{ab initio} data proving that our
interpretation of the phenomenon is adequate.
The interpretation for argon is very similar.
The model is formed by the levels~Ar$\,1s$, Ar$\,1s^{-1} 4p$, and
Ar$\,1s^{-1} 4s$.
The x~rays couple~Ar$\, 1s$ and Ar$\,1s^{-1} 4p$ whereas the laser
couples~Ar$\,1s^{-1} 4p$ and Ar$\,1s^{-1} 4s$.
The mechanism of EIT for x~rays is very similar to EIT in the optical domain,
the only difference being an appreciable decay width of the third
state~(Ar$ \, 1s^{-1} 4s$).
The transparency is therefore due to a splitting into an Autler-Townes
doublet and not primarily due to destructive interference of two
excitation pathways~\cite{Fleischhauer:ET-05}.

In Fig.~\ref{fig:XLperp}, we show the impact of laser dressing for
perpendicular laser and x-ray polarization vectors.
In contrast to the case of parallel polarization vectors in Fig.~\ref{fig:XLpar},
the suppression of x-ray absorption is appreciably smaller.
In fact, the curve for the lower intensity in the upper panel
looks nearly identical to Fig.~\ref{fig:x-ray-abs}.
The Ar$\, 1s \to 4p$~resonance in the lower panel is split into two peaks.
Also minute bumps can be perceived beyond the $K$~edge.
Let us analyze the underlying quantum optical mechanisms:
we have no direct dipole coupling of~Ar$\, 1s^{-1} 4p$ and
Ar$\, 1s^{-1} 4s$~states.
This excludes the mechanism discussed for the case of parallel polarization
vectors.
The suppression of the absorption therefore is attributed to arise
partly through a line broadening caused by the laser
dressing~\cite{Buth:ET-07}.
Additionally, the Ar$\, 1s^{-1} 4p$ and Ar$\, 1s^{-1} 3d$~states are in
resonance within the linewidth broadening due to the inner-shell decay.
The radial dipole coupling matrix element between the two states is
even larger than the coupling between the Ar$\, 1s^{-1} 4p$ and
Ar$\, 1s^{-1} 4s$~states.
We conclude that these two states together with the ground state Ar$\,1s$
form a three-level ladder-type model.
Such models also exhibit EIT---although not in a strict sense
(see Ref.~\onlinecite{Fleischhauer:ET-05} and references therein for
details)---which we observe here for the first time in the x-ray regime.

\begin{figure}
  \includegraphics[clip,width=\hsize]{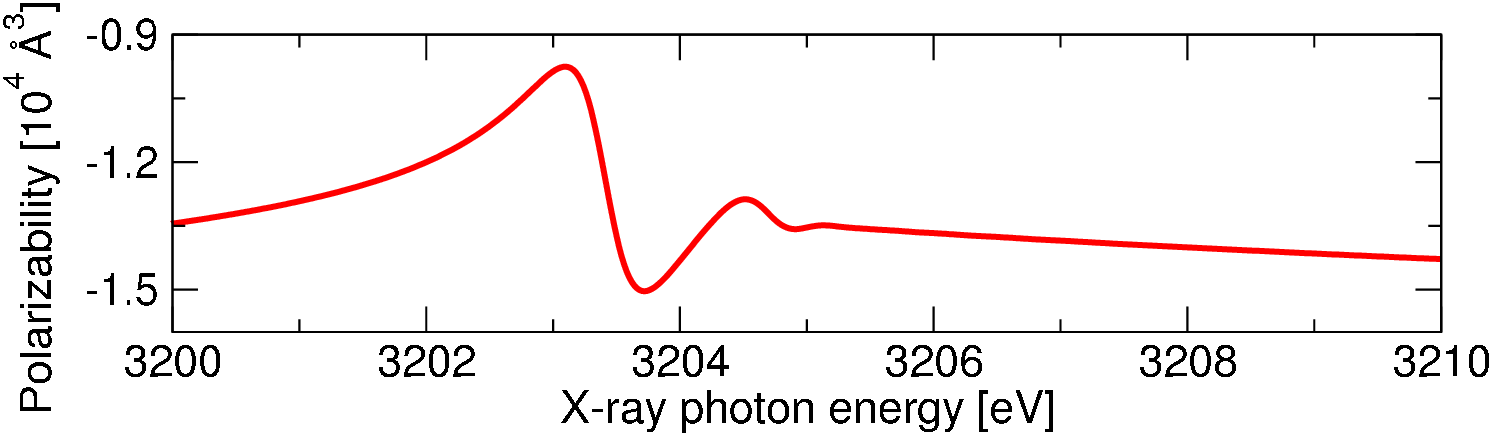}
  \caption{(Color online) X-ray polarizability of argon near the $K$~edge
           without laser dressing.}
  \label{fig:polarizability}
\end{figure}

In Fig.~\ref{fig:sigma_argon}, we show the dependence of the x-ray absorption
cross section of argon on the Ar$\,1s \to 4p$~resonance on the intensity of the
dressing laser.
The cross section without dressing laser, $128.6 \kilobarns$, is indicated.
Above a laser intensity of~$10^{11} \U{W/cm^2}$, the cross section drops
steeply towards the
value~$35.6 \kilobarns$ at~$10^{13} \U{W/cm^2}$ for parallel laser and x-ray
polarization vectors.
This corresponds to an overall reduction of the cross section with laser and
without of a factor of~$3.6$.
For perpendicular polarization vectors, only a reduction by a factor
%
%
%
%
of~$1.9$ is found.
We observe that the speed of the drop of the cross section decreases noticeably
near~$10^{13} \U{W/cm^2}$.
Eventually, the cross section seems to saturate beyond~$10^{13} \U{W/cm^2}$
before appreciable strong-field ionization of the ground-state atoms takes
place around~$10^{14} \U{W/cm^2}$.

\begin{figure}
  \begin{center}
    \includegraphics[clip,width=\hsize]{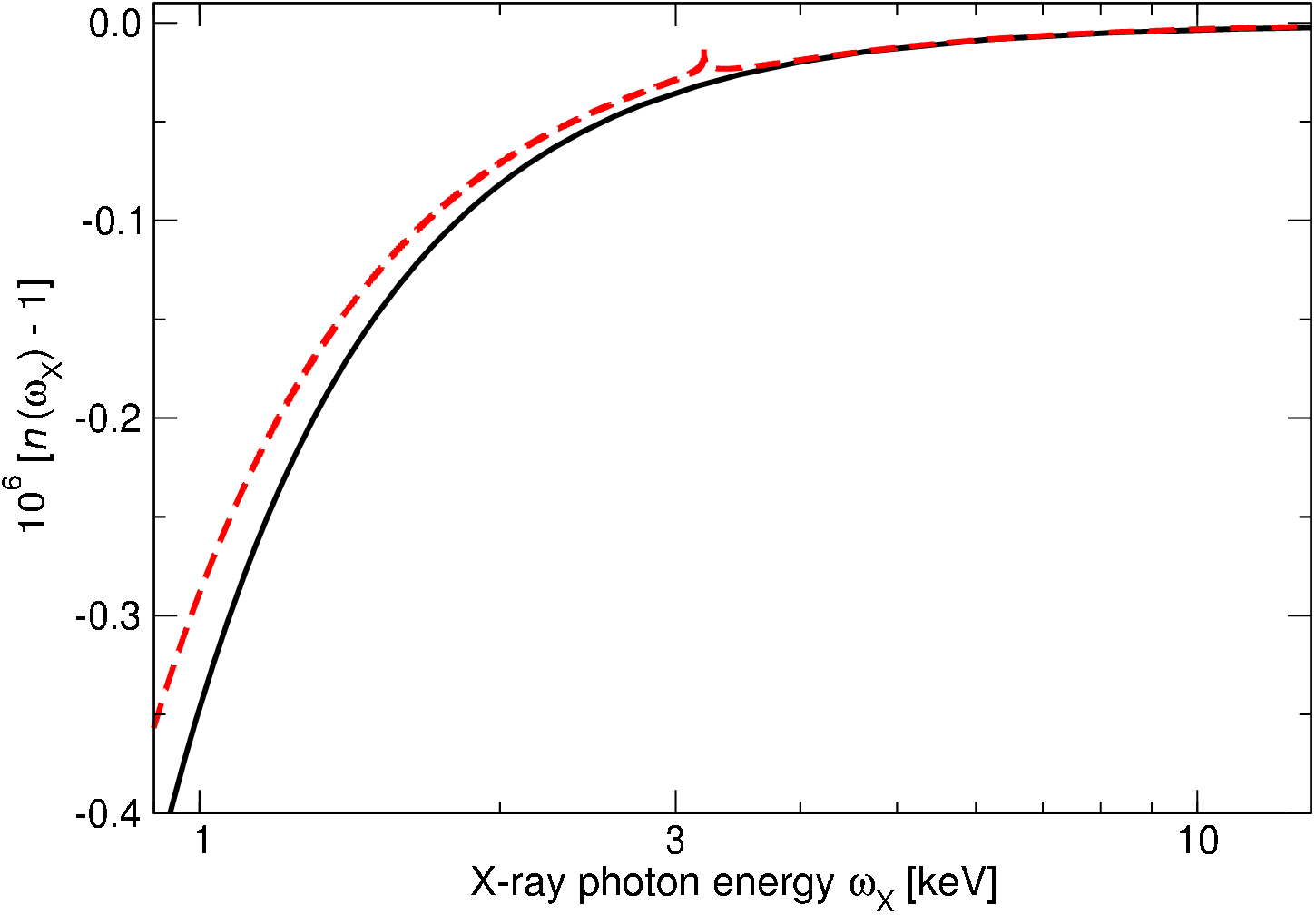}
    \caption{(Color online) Refractive index of argon without laser dressing.
             The theoretical data are represented by the dashed, red curve.
             Semiexperimental data (using the Kramers-Kronig relation to
             determine the real part of the refractive index from the
             experimental cross section data) was taken from Fig.~4 in
             Ref.~\onlinecite{Liggett:RI-68}.
             The digitized experimental curve was fitted with~$n(\omega\I{X})
             - 1 = a_0 / \omega\I{X}^{a_1}$ yielding~$a_0 = -672233 \eV^{a_1}$
             and $a_1 = 2.096$ for the two parameters.
             Then, the curve was shifted such that $n(606 \U{Ry})-1
             = -0.503 \E{8}$ (see ``This paper'' column of Table~V in
             Ref.~\onlinecite{Liggett:RI-68}).}
    \label{fig:refidx}
  \end{center}
\end{figure}

These observations can be understood in terms of the Rabi
frequency~$\Omega\I{L} = d \, E\I{L,0}$ of the laser coupling of
the Ar$\,4s$ and Ar$\,4p$~levels.
%
%
Here, $d = \bra{4s} z \ket{4p} = 3.5 \bohr$ is the dipole coupling matrix
element between the two states and $E\I{L,0} = \sqrt{8 \pi \, \alpha \,
I_{X,0}}$ the peak electric field.
For the laser intensities~$10^{11} \U{W/cm^2}$, $10^{12} \U{W/cm^2}$, and
$10^{13} \U{W/cm^2}$, we find the
%
%
%
%
%
Rabi frequencies~$\frac{1}{26 \U{fs}}$, $\frac{1}{8 \U{fs}}$, and
$\frac{1}{2.6 \U{fs}}$.
The frequency~$\Omega\I{L}$ should be comparable with the inverse core-hole
lifetime of argon
%
%
of~$1 \U{fs}$~\cite{Campbell:WA-01} such that it beats the decay and
absorption is suppressed.
The $\Omega\I{L}$ also determines the magnitude of the splitting of the
lines in the Autler-Townes doublet~\cite{Buth:ET-07}.
However, the increasing laser intensity also leads to a broadening of
the Rydberg states due to ionization.
The line broadening competes with the Rabi flopping and finally slows down
the drop of the cross section with increasing laser intensity in
Fig.~\ref{fig:sigma_argon}.
It potentially leads to a halt of the drop before ionization kicks in.

The real part of the energy shift in the x-ray field determines the
atomic polarizability~(\ref{eq:polartot}).
It is plotted for argon around the $K$~edge in Fig.~\ref{fig:polarizability}.
Clearly the dispersion curve of the Ar$\,1s \to 4p$~pre-edge resonance is
perceivable.

From the atomic polarizability, we determine the refractive index at normal
temperature and pressure~\nocite{Liggett:RI-68,Kingston:RI-64}\cite{footnote4}
from Eq.~(\ref{eq:refractive}).
In Fig.~\ref{fig:refidx}, we compare our theoretical results with the data from
Fig.~4 of Liggett and Levinger~\cite{Liggett:RI-68}.
We find a very good agreement above the $K$~edge of argon.
Below $300 \eV$, the refractive index is not even qualitatively reproduced
because our approximation breaks down that the refractive index is determined
only by $K$-shell electrons;
the electrons in higher lying shells become significant.

\begin{figure}
  \includegraphics[clip,width=\hsize]{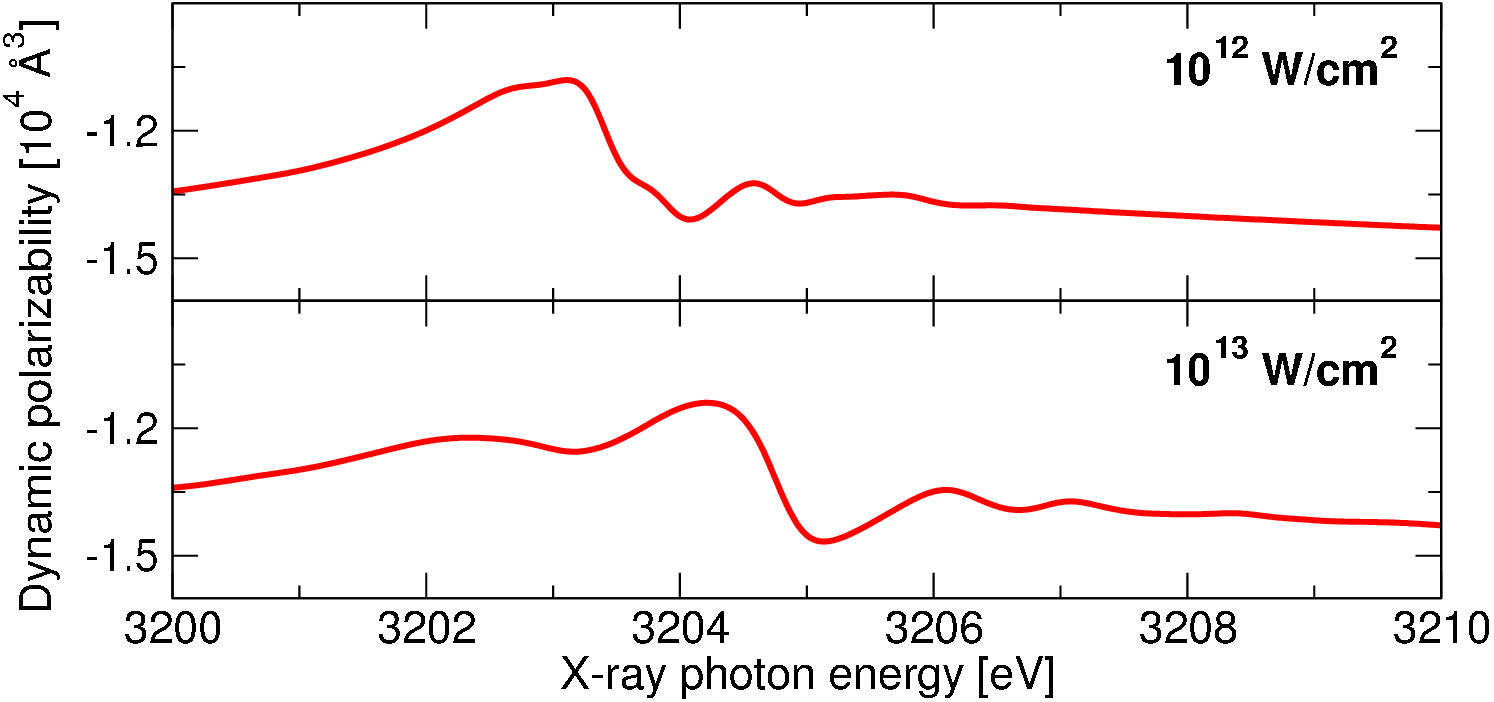}
  \caption{(Color online) X-ray polarizability of argon for parallel
           laser and x-ray polarization vectors.
           In the upper panel, the intensity of the dressing laser
           is~$10^{12} \U{W/cm^2}$;
           it is $10^{13} \U{W/cm^2}$ in the lower panel.}
  \label{fig:XLparPol}
\end{figure}

\begin{figure}
  \includegraphics[clip,width=\hsize]{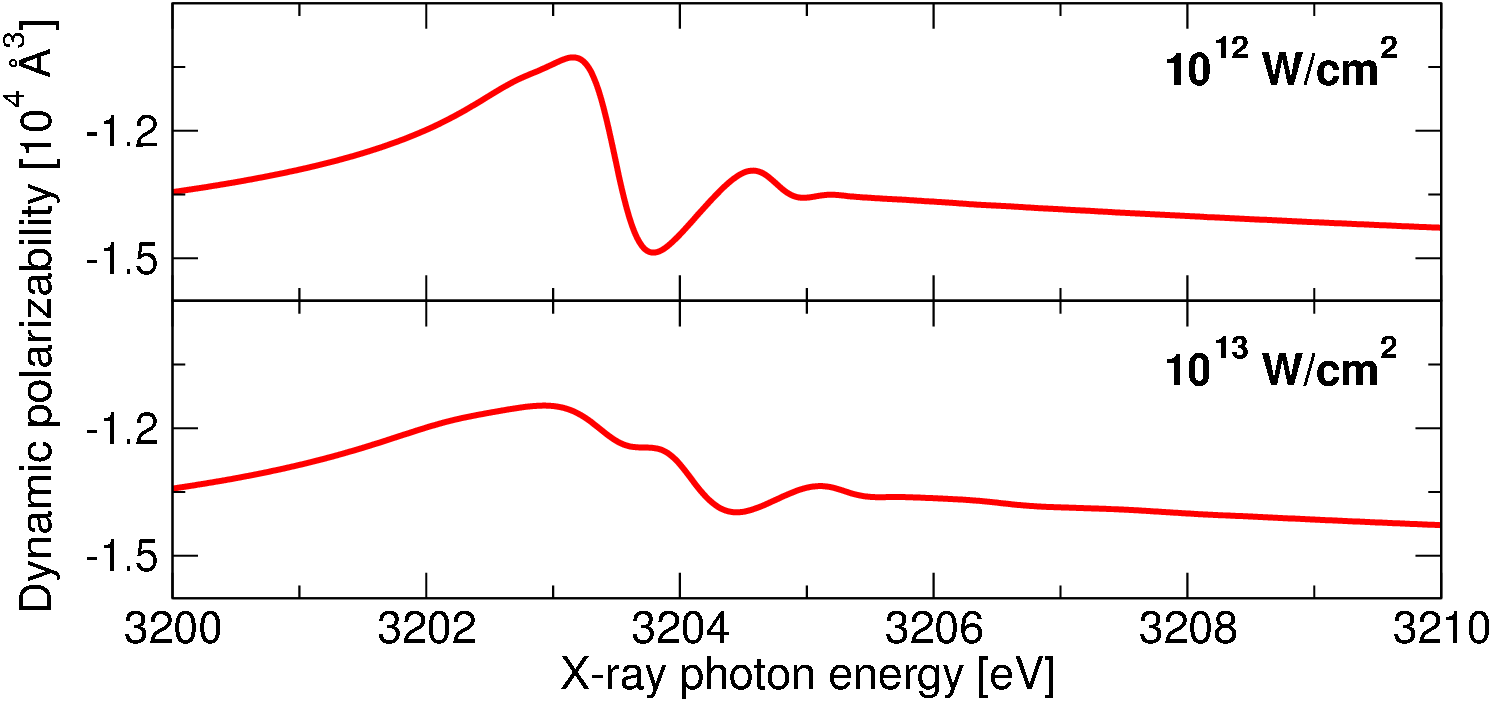}
  \caption{(Color online) X-ray polarizability of argon for
           perpendicular laser and x-ray polarization vectors.
           The dressing laser has an intensity of~$10^{12} \U{W/cm^2}$ in the
           upper panel and of~$10^{13} \U{W/cm^2}$ in the lower panel.}
  \label{fig:XLperpPol}
\end{figure}

The atomic polarizability and thus the refractive index is changed by laser
dressing.
In Fig.~\ref{fig:XLparPol}, we investigate the impact of parallel laser
and x-ray polarization vectors and in Fig.~\ref{fig:XLperpPol}, we
plot the dynamic polarizability for perpendicular polarization vectors.
Generally, the polarization---and thus the refractive
index~(\ref{eq:refractive})---is dominated by the~$\vec A\I{X}^2$ term.
Hence the impact of the laser dressing is expected to be small because
it is mediated by the $\hat{\vec p} \mul \vec A\I{X}$~term.
This expectation is confirmed by Figs.~\ref{fig:XLparPol} and
\ref{fig:XLperpPol};
the magnitude of refraction and dispersion of x~rays is small with
or without laser dressing.
The shape of the curves in the two figures is, however, very different.
While for parallel polarization vectors a substantial change of the
shapes is obtained for~$10^{12} \U{W/cm^2}$, the corresponding curve for
perpendicular polarizations is nearly unchanged compared with
Fig.~\ref{fig:polarizability}.
The different behavior again reveals the different underlying quantum
optical process already discussed for absorption.

\begin{figure}
  \includegraphics[clip,width=\hsize]{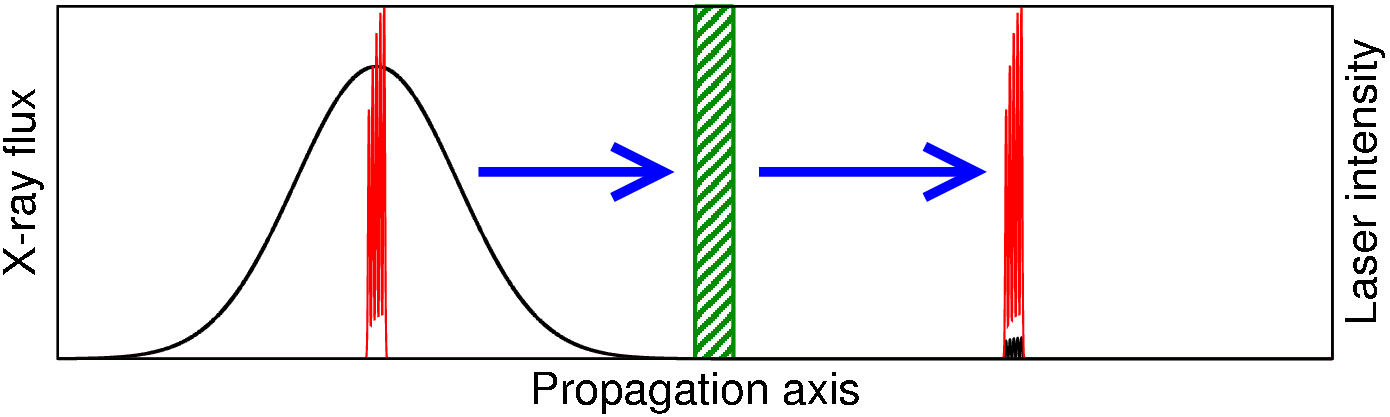}
  \caption{(Color online) X-ray pulse shaping with electromagnetically induced
           transparency.
           Propagation of a laser pulse [Eq.~(\ref{eq:lasshape})] (red, high, narrow)
           and an x-ray pulse [Eq.~(\ref{eq:XrayGaussEnv})]
           (black, Gaussian, broad) on the right hand side through a cell
           (green, hatched) filled with argon gas.
           The pulses copropagate from left to right with the speed of light.}
  \label{fig:ppset}
\end{figure}

In Fig.~\ref{fig:ppset}, we demonstrate the principle of x-ray pulse shaping
based on the effect electromagnetically induced transparency for x~rays.
On the left hand side of the gas cell, the initial laser and x-ray
pulses are shown.
The laser pulse has the shape of Eq.~(\ref{eq:lasshape});
the x-ray pulse has a Gaussian envelope~(\ref{eq:XrayGaussEnv}).
The pulses copropagate from left to right with the speed of light
in vacuum~$c$ through the gas cell which is filled with argon.
Also in the cell the propagation speed deviates only minutely from~$c$
(see the appendix).
After passing through the cell---the right hand side of the figure---a pulse
shape similar to the shape of the laser pulse is cut
out of the broad x-ray pulse;
the laser pulse remains unchanged.
The fairly elaborate pulse shape~(\ref{eq:lasshape}) was imprinted approximately
on the x-ray pulse.
In Fig.~\ref{fig:ppshape}, we show a close-up of the x-ray pulse on the
right hand side of Fig.~\ref{fig:ppset} for two different intensities of
the dressing laser.
We see a pronounced reduction of the initial peak x-ray flux~$J\I{X,0}$
[Eq.~(\ref{eq:XrayGaussEnv})].
Due to the nonlinear dependence of the x-ray absorption cross section
on the laser intensity (Fig.~\ref{fig:sigma_argon}), the resulting
transmitted pulse varies not only in the flux but also the general
shape differs somewhat from the initial pulse.
Overall the peak x-ray flux drops for~$10^{13} \U{W/cm^2}$ by a factor of~10.
For neon, we found a stronger transmission on resonance which leads to
a suppression of only a half for the same peak laser
intensity~\cite{Buth:ET-07}.

Only the flux envelope of the x-ray pulse has been modified in
Fig.~\ref{fig:ppshape}.
The time evolution of the phase of the x-ray pulses remains random.
With the EIT for x~rays method, the phase cannot be effectively manipulated
because the magnitude of the polarizability and variation of the
polarizability due to laser dressing are very small.
Hence the magnitude and variation of the index of refraction of the gas is
also very small.
The x~rays are basically fully absorbed before the phase can be influenced
noticeably.
This finding is in contrast to EIT for optical wavelengths where enormous
changes of refraction and dispersion of the medium on the resonance
can be observed and exploited see, \eg, Refs.~\onlinecite{Hau:SL-99,%
Phillips:SL-01,Lukin:ET-01,Fleischhauer:ET-05}.

\section{Conclusion}
\label{sec:conclusion}

In this paper, we investigated the complex index of refraction of atoms
in the light of an intense optical laser in the x-ray region.
This is a so-called two-color problem.
The laser intensity is assumed to be sufficiently low such that
the atoms are only dressed but not excited or ionized.
We devised an \emph{ab initio} theory to compute the basic atomic quantities,
the dynamic dipole polarizability and the photoabsorption cross section,
from which the real and imaginary part of the index of refraction,
respectively, are determined.
We describe the fundamental interaction of the atoms with the light in
terms of quantum electrodynamics.
The determination of the atomic properties involves resolvents.
We use a single-vector Lanczos algorithm to compute the involved
resolvents directly.
The index of refraction is a concept of classical electrodynamics;
it follows from the Maxwell equations.
The transition from the quantum mechanics to classical physics
is made by expressing the classical macroscopic polarizability in terms of
quantum mechanical atomic polarizability and atomic absorption cross section.
Finally, we consider a laser and an x-ray pulse copropagating through a
gas cell using rate equations.

Our theory is applied to argon.
We study the x-ray polarizability and absorption over a large range of
dressing-laser intensities for parallel and perpendicular laser and
x-ray polarization vectors.
For parallel polarizations, we find electromagnetically induced
transparency~(EIT) for x~rays on the Ar$\, 1s \to 4p$~pre-edge resonance.
The absorption is suppressed by a factor
of~$3.6$ when the laser is present.
Also for perpendicular polarizations, the cross section on resonance shows
a noticeable drop by a factor of~$1.9$ for a laser intensity
of~$10^{13} \U{W/cm^2}$.
As an application of EIT for x~rays, the control of the absorption of the x~rays
allows one to imprint the shape of the laser pulse on x~rays.
We show the transmitted x-ray pulses for two different laser intensities
which vary, apart from the flux, also due to the nonlinear dependence
of the cross section on the laser intensity.

Our work opens up numerous possibilities for future research.
The $K$~edge of argon is at~$3205.9 \eV$~\cite{Thompson:XR-01} which
implies a wavelength
of the x~rays of~$3.9 \angstrom$, \ie, our work opens up possibilities
to shape hard x-ray pulses.
Ultrashort, femtosecond, hard x-ray pulses come into reach.
Especially, in conjunction with the emerging x-ray free electron lasers,
our results may prove useful.
The ability of pulse shaping x~rays opens up possibilities for all-x-ray
pump-probe experiments and quantum control of inner-shell processes.

\begin{acknowledgments}
We thank Linda Young for fruitful discussions.
C.B.~was partly funded by a Feodor Lynen Research Fellowship from
the Alexander von Humboldt Foundation.
This work was supported by the Office of Basic Energy Sciences, Office of
Science, U.S.~Department of Energy, under Contract No.~DE-AC02-06CH11357.
\end{acknowledgments}

\appendix*
\section{Secondary physical processes}

\begin{figure}
  \includegraphics[clip,width=\hsize]{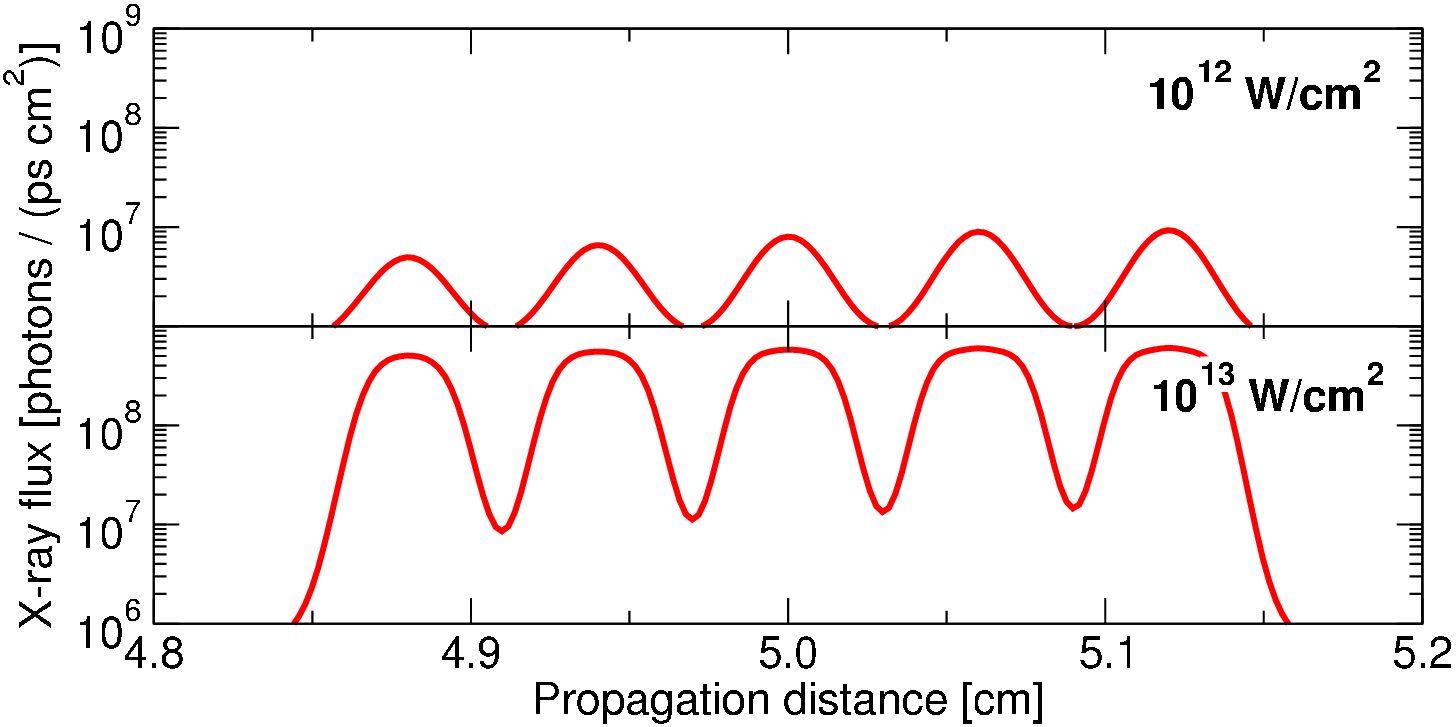}
  \caption{(Color online) The x-ray pulse shapes that are obtained in argon gas
           with the setup in Fig~\ref{fig:ppset} for two regimes of
           laser intensity: $10^{12} \U{W/cm^2}$ (upper panel) and
           $10^{13} \U{W/cm^2}$ (lower panel).}
  \label{fig:ppshape}
\end{figure}

%
%
\begin{table*}
  \centering
  \begin{ruledtabular}
    \begin{tabular}{ccccc}
     $\omega$ [$\eV$] & $v\I{p}/c - 1$ & $v\I{g}/c - 1$ & $v\I{p,pl}/c - 1$
       & $v\I{g,pl}/c - 1$ \\
     \hline
     \phantom{000}1.55 & $-1.3 \E{-3}$  & $-1.3 \E{-3}$ & $1.4
       \E{-8\phantom{0}}$ & $-1.4 \E{-8\phantom{0}}$ \\
     3203.42 & $\phantom{-}9.6 \E{-8}$ & $\phantom{-}4.0 \E{-4}$
       & $3.4 \E{-15}$ & $-3.4 \E{-15}$
    \end{tabular}
  \end{ruledtabular}
  \caption{Estimate of the changes of the phase and group velocity due to
           secondary physical effects.
           We consider refraction and dispersion of the laser and the x~rays
           by the argon-gas medium and the plasma for $I\I{L,0}
           = 10^{13} \U{W/cm^2}$.}
  \label{tab:secphys}
\end{table*}

Numerous secondary physical processes occur in course of the pulse propagation.
The absorption of x~rays leads to the production of ions and free electrons;
in other words, a plasma.
The ion production is governed by the rate equation
\begin{equation}
  \Delta n'_{\#}(z_i) = n'_{\#}(z_i) \, \sigma'(z_i) \, J'\I{X}(z_i) \,
    \Delta t \; .
\end{equation}
In the inner part of the gas cell the ion number density is constant;
its maximum
is~$1.3 \E{13} \U{cm^{-3}}$.
The ions which are created by photoionization have a $K$~vacancy.
Each resonantly absorbed x~ray generates a decay cascade which leads to the
emission of~$3.8$~electrons (Fig.~1(b) in Ref.~\onlinecite{Doppelfeld:CL-93}).

The phase velocity~\cite{Jackson:CE-98} of laser and x-ray pulses in the medium is
\begin{equation}
  \label{eq:vphase}
  v\I{p}(\omega) = \frac{c}{n(\omega)}
\end{equation}
and their group velocity is~\cite{Jackson:CE-98,Fleischhauer:ET-05}:
\begin{equation}
  \label{eq:groupvel}
  v\I{g}(\omega) = \frac{c}{n(\omega) + \omega \frac{\differential n}
    {\differential \omega}} \; ,
\end{equation}
where $n(\omega)$ denotes the real part of the refractive index
in this appendix.
We will show that both velocities differ only minutely from the corresponding
velocities in vacuum for our choice of parameters.

The index of refraction of a plasma is
\begin{equation}
  n\I{pl}(\omega) = \sqrt{1 - \frac{\omega\I{p}^2}{\omega^2}} \; ,
\end{equation}
with the plasma frequency~$\omega\I{p} = \sqrt{4 \pi \,
n\I{e}}$~\cite{Jackson:CE-98}.
The group velocity for a plasma is
\begin{equation}
 \label{eq:vgrpplsm}
  v\I{g,pl}(\omega) = c \, n\I{pl}(\omega) \; .
\end{equation}
%

%
%
%
The refractive index of argon gas for laser light is obtained from
Eq.~(\ref{eq:refractive}).
We use the dynamic polarizability of argon atoms for~$\omega\I{L}
= 1.55 \eV$~\cite{Rahman:DD-90,footnote5}
which is~$11.42 \, e^2 \, a_0^2 \Hartree^{-1}$.
The real part of the refractive index
%
%
follows with Eq.~(\ref{eq:refractive}) to~$n(1.55 \eV) = 1.0013$
for the physical parameters from Sec.~\ref{sec:compdet}.

The group velocity of x~rays in argon gas follows from Eq.~(\ref{eq:groupvel}).
We insert the derivative of the index of refraction letting~$\mu = 0$ in
Eq.~(\ref{eq:refractive}) yielding
\begin{equation}
  \label{eq:vgrpxrays}
  v\I{g,X}(\omega\I{X}) = \frac{c}{1 + 2 \pi \, n_{\#} \, [\alpha\I{I}
    (\omega\I{X}) + \omega\I{X} \, \frac{\differential \alpha\I{I}}
    {\differential \omega}(\omega\I{X})]} \; .
\end{equation}
The polarizability of argon on resonance is taken from the curve in
Fig.~\ref{fig:polarizability} and its numerical derivative.
The group velocity of the laser in argon gas~$v\I{g,L}(\omega\I{L})$ is
obtained from Eq.~(\ref{eq:vgrpxrays})---with ``X'' replaced by~``L''---by
inserting the slope of the line used for the linear interpolation of
the dynamic polarizability~\cite{Rahman:DD-90,footnote5}.

In Table~\ref{tab:secphys}, we list exemplary estimates of the secondary
physical effects.
We assume a laser intensity of~$10^{13} \U{W/cm^2}$.
Overall we observe that these effects are very small under the conditions of
this paper.
However, for increasing gas densities and/or increasing laser intensities, these
factors become relevant and have to be accounted for.
The largest difference to~$c$ occurs for the phase and group velocities of the
laser in the argon gas.
We observe that the maximum of the phase velocities and the group velocities
are essentially the same.
This is unlike EIT for optical wavelengths where the group velocity and
phase velocity on resonance differ largely, leading to slow
light~\cite{Hau:SL-99,Fleischhauer:ET-05}.
For lower laser intensities, the plasma related velocities, $v\I{p,pl}$ and
$v\I{g,pl}$
increase slightly because a larger fraction of the x-ray pulse is
absorbed leading to a denser plasma.
Conversely, these velocities are smaller for higher laser intensities as long as
the laser intensity remains far below the saturation intensity.
As soon as the laser begins to contribute to the plasma formation, $v\I{p,pl}$
and $v\I{g,pl}$ deviate noticeably from~$c$.


\begin{thebibliography}{63}
\bibitem[{foo({\natexlab{a}})}]{footnote2}
\bibinfo{note}{We assume that the continuum of core-excited states is
  discretized. In Sec.~\ref{sec:HFSapprox} we introduce a complex absorbing
  potential and a basis set to accomplish this.}
\bibitem[{foo({\natexlab{b}})}]{footnote1}
\bibinfo{note}{Equation~(\ref{eq:Hintlight}) was obtained by converting the
  equation from SI~units to atomic units. Therefore factors~$\alpha$, the
  fine-structure constant, are missing which are typically found when
  converting equations from Gaussian units to atomic units.}
\bibitem[{foo({\natexlab{c}})}]{footnote3}
\bibinfo{note}{We restrict ourselves to single core holes due to only moderate
  x-ray intensities.}
\bibitem[{foo({\natexlab{d}})}]{footnote4}
\bibinfo{note}{The temperature and pressure of the argon gas, for which the
  refractive index is determined, are not stated in
  Ref.~\onlinecite{Liggett:RI-68}. Yet in Ref.~\onlinecite{Liggett:RI-68}, the
  authors refer to a similar study by Kingston~\cite{Kingston:RI-64} who gives
  results for normal temperature and pressure~(NTP), \ie, 
  $20\Celsius$ and $1 \atm$. Hence, to compare with Liggett and Levinger, we
  assume NTP for argon gas.}
\bibitem[{foo({\natexlab{e}})}]{footnote5}
\bibinfo{note}{The value of~$\omega\I{L} = 1.55 \eV = 0.057 \Hartree$ is
  determined by linear interpolation between the points~$0.05 \Hartree$ and
  $0.1 \Hartree$ from Ref.~\onlinecite{Rahman:DD-90}.}
\end{thebibliography}
\end{document}